\documentclass[11pt]{article}
\usepackage{a4}
\usepackage{amsmath}
\usepackage{amssymb}
\usepackage[dvips]{graphicx}
\usepackage{subfigure}
\def\beq{\begin{equation}}
\def\eeq{\end{equation}}
\def\beqa{\begin{eqnarray}}
\def\eeqa{\end{eqnarray}}
\def\n{\nonumber \\}

\def\e{{\,\rm e}\,}

\newcommand {\tr}{{\rm tr}\,}

\def\dag{\dagger}
\newcommand{\id}{{1\!\!1}} 

\begin{document}

\vspace*{1.0cm}
\begin{flushright}
{SAGA-HE-279}
\end{flushright}
\vskip 1.0cm

\begin{center}
{\large{\bf Probability distribution over some phenomenological models\\
 in the matrix model compactified on a torus}}
\vskip 1.0cm

{\large Hajime Aoki\footnote{e-mail
 address: haoki@cc.saga-u.ac.jp}
}
\vskip 0.5cm

{\it Department of Physics, Saga University, Saga 840-8502,
Japan  }\\

\end{center}

\vskip 1cm
\begin{center}
\begin{bf}
Abstract
\end{bf}
\end{center}
We study some phenomenological models in a matrix model
corresponding to the IIB matrix model compactified on a six-dimensional torus 
with magnetic fluxes.
Extending our previous works,
we examine a wider class of models:
a Pati-Salam-like model with a gauge group 
${\rm U}(4) \times {\rm U}_L(2) \times {\rm U}_R(2)$,
and models where the gauge group U(4) is broken down to 
${\rm U}_c(3) \times {\rm U}(1)$
and/or ${\rm U}_R(2)$ is broken down to ${\rm U}(1)^2$.
We find all the matrix configurations that yield matter content of 
all the phenomenological models whose gauge group is a subgroup of U(8).
We then estimate semiclassically a probability distribution for the appearance
of the phenomenological models.

\newpage
\setcounter{footnote}{0}
\section{Introduction}
\setcounter{equation}{0}

The standard model (SM) of particle physics agrees well with experiments.
While phenomenological models beyond the SM will be explored
in (near-)future experiments,
some theoretical guides may be helpful.
It may also be important to reconsider why the SM is so successful
and what we should ask nature next.

On the other hand, the SM is unsatisfactory as a final theory,
and the string theory is expected to be an ultimate theory
including gravity.
Phenomenological models inspired by string theories
have been studied extensively
(see, for instance, ref.~\cite{Ibanez:2012zz, Blumenhagen:2006ci}).
However, a serious problem in the string theory 
is that it has too many vacua.

A candidate for a nonperturbative formulation of 
string theory is 
the matrix model (MM) 
\cite{Banks:1996vh}-\cite{Dijkgraaf:1997vv}.
Since the MM has a definite action and measure,
we can, in principle, dynamically compare the string vacua, and 
calculate everything, such as
spacetime dimensions, gauge groups, and matter contents.
Indeed, spacetime structures have been analyzed intensively,
and four-dimensionality seems to be preferred 
in the IIB matrix model 
\cite{Aoki:1998vn}-\cite{Kim:2012mw}.
Then,
assuming that our spacetime is obtained, 
we will consider how the SM and some phenomenological models appear from the MM,
and estimate the probability distribution of their appearance.
Such phenomenological studies in the MM may give us a guide
for exploring phenomenological models beyond the SM.

An important ingredient of the SM is the chirality of fermions.
We usually obtain a chiral spectrum on our spacetime 
by introducing a nontrivial topology in the extra dimensions:
Euler characteristics of compactified manifolds,
special boundary conditions at orbifold singularities,
the intersection numbers of D-branes,
etc., give nontrivial topologies.
Also from the MM,
chiral fermions and the SM matter content have been obtained 
by considering toroidal compactifications with magnetic fluxes \cite{Aoki:2010gv,Aoki:2012ei}
and intersecting D-branes \cite{Chatzistavrakidis:2011gs}\footnote{
Studies based on fuzzy spheres are given in refs.
\cite{AIMN}-\cite{Aoki:2010hx}.
MMs for orbifolds and orientifolds are studied in refs.
\cite{Aoki:2002jt}-\cite{Itoyama:1998et}. 
}.
The former is similar to magnetized D-branes wrapping on a torus,
which are T-dual to the latter.

In this paper, we will study phenomenological models in the MM compactified on a torus,
extending our previous works \cite{Aoki:2010gv,Aoki:2012ei}.
We here examine a wider class of phenomenological models than in ref.~\cite{Aoki:2012ei}.
In order to embed the SM fermions with three generations into matrices
in our formulation,
the gauge group must be a subgroup of U(8) or larger groups.
We then find all the matrix configurations that yield 
all the phenomenological models
whose gauge group is a subgroup of U(8).
Exhausting all the solutions is necessary for studying a semiclassical analysis.

We here note that
even if a matter field is massless at the tree level in the MM,
which might be interpreted as phenomena at the Planck scale,
it would obtain a mass through quantum corrections at low energies.
While masses of the gauge field and fermionic field can be protected
by the gauge and the chiral symmetry, respectively,
a mass of a scalar field is difficult to protect,
which is well known as the naturalness or the hierarchy problem.
Thus, although the matrices provide scalar fields
with the same representation under the gauge group as the Higgs field,
it is difficult to keep them massless.
Then, we will first find matrix configurations
that provide the gauge and the fermionic fields,
without considering how those Higgs candidates remain massless.
We will next assume situations where 
the Higgs mass is protected by the supersymmetry,
and find matrix configurations that provide the Higgsino fields as well.

We then study the dynamics of the MM semiclassically,
and estimate a probability distribution for the appearance
of the phenomenological models\footnote{
Probability distributions over the string landscape \cite{Susskind:2003kw}
have been estimated, for instance, by
counting the number of flux vacua \cite{Denef:2004ze},
and by considering cosmological evolutions \cite{Bousso:2000xa,Bousso:2006nx}.
}.
The fact that we could perform these analyses is 
an advantage of the MM\footnote{
A related work is given in the MM of noncritical strings \cite{Chan:2012ud}.}.
On the other hand,
there remain some important issues in the formulations of MMs,
such as the relations between MMs and string theory,
interpretations of spacetime and matter in the matrices,
and how to take large-$N$ limits.
Reversely, we can expect that these phenomenological studies may
give some hints for those problems in MMs.

In section~\ref{sec:model},
we briefly review a formulation of topological configurations on a torus;
a similar review was given in ref.~\cite{Aoki:2012ei}.
We then find matrix configurations that provide phenomenological models
without the Higgsino fields
in section~\ref{sec:confpheno},
and those with the Higgsino fields in section~\ref{sec:conhiggs}.
In section \ref{sec:probpheno}, we estimate the probability distribution
over the phenomenological models.
Section \ref{sec:conclusion} is devoted to the conclusions and discussions.
In appendix \ref{sec:q}, detailed calculations for determining the fluxes
are shown.

\section{Review of topological configurations on a torus}
\label{sec:model}
\setcounter{equation}{0}

We begin with a brief review of the IIB MM \cite{IKKT,Aoki:1998bq}.
Its action has a simple form
\beq
S_{\rm IIBMM}  = -{1\over g^2_{\rm IIBMM}}  ~\tr
\left({1\over 4}[A_{M},A_{N}][A^{M},A^{N}]
+{1\over 2}\bar{\Psi}\Gamma ^{M}[A_{M},\Psi ]\right) \ ,
\label{IIBMMaction}
\eeq
where $A_{M}$ and $\Psi$ are $N \times N$ Hermitian matrices.
They are also a ten-dimensional vector and a Majorana-Weyl spinor, respectively.
Performing a kind of functional integration
as a statistical system, and taking a suitable 
large-$N$ limit, one can obtain a
nonperturbative formulation of string theory.
Since the measure as well as the action is defined definitely,
we can calculate everything in principle.
Another notable feature is that both spacetime and matter 
emerge from the matrices.
Spacetime can be interpreted as an eigenvalue distribution of the bosonic matrices $A_M$,
and its structures have been analyzed dynamically \cite{Aoki:1998vn}-\cite{Kim:2012mw}.
Arguments for local fields on it have been given 
\cite{Iso:1999xs,Nishimura:2012rs}.
It has also been shown that noncommutative (NC) space and matter fields on it are
described in the MM rather elegantly \cite{Connes:1997cr,Aoki:1999vr}\footnote{
It has also been shown that curved spaces can be described by 
interpreting the matrices as differential operators \cite{Hanada:2005vr}.}.
 
We then consider toroidal compactifications of $M^4 \times T^6$
with $T^6$ carrying magnetic fluxes.
Toroidal compactifications have been studied in
Hermitian matrices \cite{Connes:1997cr,Taylor:1996ik}
and in unitary matrices \cite{Polychronakos:1997fw}-\cite{Tada:1999mm}.
We here use a finite-unitary-matrix formulation for NC tori.
It is defined by the twisted Eguchi-Kawai model 
\cite{Eguchi:1982nm}-\cite{GonzalezArroyo:1982hz}
(see, for instance, refs.~\cite{Ambjorn:1999ts}-\cite{Ambjorn:2000cs}).
One may consider background configurations corresponding to
\beqa
A_\mu &\sim& x_\mu \otimes \id \ , \n
e^{i A_i} &\sim& \id \otimes V_i \ ,
\label{Areldec}
\eeqa
with $\mu=0,\ldots,3$ and $i=4,\ldots,9$,
where $A_M$ stand for the Hermitian matrices in the IIB MM (\ref{IIBMMaction}).
Unitary matrices $V_i$ represent $T^6$, while
$x_\mu$ represent our spacetime $M^4$.
One can also consider situations where our spacetime is compactified as well.
A ten-dimensional NC torus with an anisotropy of sizes
between four and six dimensions can be described by a
unitary MM
\beqa
S_{b} &=& -\beta {\cal N} \, \sum_{i \ne j} 
{\cal  Z}_{ji}
~\tr ~\Bigl({\cal V}_i\,{\cal V}_j\,{\cal V}_i^\dag\,{\cal V}_j^\dag\Bigr) 
-\beta' {\cal N} \, \sum_{\mu \ne \nu} 
{\cal  Z}_{\nu\mu}
~\tr ~\Bigl({\cal V}_\mu\,{\cal V}_\nu\,{\cal V}_\mu^\dag\,{\cal V}_\nu^\dag\Bigr) \n
&&-\beta'' {\cal N} \, \sum_{i \mu} \left[
{\cal  Z}_{\mu i}
~\tr ~\Bigl({\cal V}_i\,{\cal V}_\mu\,{\cal V}_i^\dag\,{\cal V}_\mu^\dag\Bigr)
+{\cal  Z}_{i \mu}
~\tr ~\Bigl({\cal V}_\mu\,{\cal V}_i\,{\cal V}_\mu^\dag\,{\cal V}_i^\dag\Bigr) \right] \ ,
\label{TEK-action}
\eeqa
where ${\cal V}_\mu$ and ${\cal V}_i$ are ${\rm U}({\cal N})$ matrices
with  $\mu, \nu =0, \ldots, 3$ and $i,j = 4, \ldots, 9$.
$\beta$, $\beta'$ and $\beta''$ are coupling constants, 
and ${\cal  Z}_{MN}$ are phase parameters called twists.
A setting of these parameters will be discussed 
in section \ref{sec:InstantonAction}.
A relation between the IIB MM (\ref{IIBMMaction}) and the
unitary MM (\ref{TEK-action}) will be argued 
in section \ref{sec:largeNprescription}.
We then consider background configurations as
\beqa
e^{i A_\mu} &\sim& {\cal V}_\mu = V_\mu \otimes \id  \ , \n
e^{i A_i} &\sim& {\cal V}_i = \id \otimes V_i \ .
\label{relcalVV1}
\eeqa
We will hereafter study the extra-dimensional space $T^6$
in the unitary MM (\ref{TEK-action}).

We then focus on $V_i$ in (\ref{relcalVV1}), 
i.e., NC $T^6$ with nontrivial topologies.
It is known that
nontrivial topological sectors are defined by the so-called 
modules in NC geometries
(see, for instance, ref.~\cite{Szabo:2001kg}).
In the MM formulations, the modules are defined
by imposing twisted boundary conditions on the matrices
\cite{Ambjorn:1999ts}-\cite{Ambjorn:2000cs},\cite{Aoki:2008ik}.
In fact, each theory with twisted boundary conditions yields a single topological sector
specified by the boundary conditions
\cite{Paniak:2002fi}-\cite{Aoki:2009fs},
while in ordinary gauge theories on commutative spaces,
a theory, for instance, with periodic boundary conditions, 
provides all the topological sectors.
However, since we now want to derive everything from a definite MM,
those topological features of NC gauge theories are not desirable.
We then introduce nontrivial topological sectors by background matrix configurations,
not by imposing twisted boundary conditions by hand.
Nontrivial topologies can be given by block-diagonal matrices \cite{Aoki:2010gv}.
We consider the following configurations:
\beqa
V_{3+j} &=& 
\begin{pmatrix}
 \Gamma_{1,j}^1 \otimes \id_{n^1_2} \otimes \id_{n^1_3} \otimes \id_{p^1}&&\cr
& \ddots& \cr
&& \Gamma_{1,j}^h \otimes \id_{n^h_2} \otimes \id_{n^h_3}\otimes \id_{p^h}
\end{pmatrix} \ , \n
V_{5+j} &=& 
\begin{pmatrix}
\id_{n^1_1} \otimes  \Gamma_{2,j}^1 \otimes \id_{n^1_3} \otimes \id_{p^1}&&\cr
& \ddots& \cr
&& \id_{n^h_1} \otimes \Gamma_{2,j}^h \otimes \id_{n^h_3}\otimes \id_{p^h}
\end{pmatrix} \ , \n
V_{7+j} &=&
\begin{pmatrix}
\id_{n^1_1} \otimes \id_{n^1_2} \otimes \Gamma_{3,j}^1  \otimes \id_{p^1}&&\cr
& \ddots& \cr
&& \id_{n^h_1} \otimes \id_{n^h_2} \otimes \Gamma_{3,j}^h \otimes \id_{p^h}
\end{pmatrix} \ , \n
\label{conf_V6}
\eeqa
with $j=1,2$.
The number of blocks is denoted by $h$.
Each block is a tensor product of four factors.
The first three factors each represent $T^2$ in $T^6=T^2 \times T^2 \times T^2$,
and the last factor provides a gauge group structure.
The configuration (\ref{conf_V6}) gives the gauge group
${\rm U}(p^1) \times {\rm U}(p^2) \times \cdots \times {\rm U}(p^h)$.

The matrices $\Gamma_{l,j}^a$ with $a=1,\ldots,h$ and $l=1,2,3$ 
in (\ref{conf_V6}) are defined by using the Morita equivalence.
For details, see, for instance, 
refs.~\cite{Aoki:2010gv},\cite{Ambjorn:1999ts}-\cite{Aoki:2008ik}.
$\Gamma_{l,j}^a$ are ${\rm U}(n^a_l)$ matrices
that satisfy the 't Hooft-Weyl algebra
\beq
\Gamma_{l,1}^a \Gamma_{l,2}^a = 
\e^{-2\pi i\frac{m^a_l}{n^a_l}}\Gamma_{l,2}^a \Gamma_{l,1}^a \ ,
\eeq
where the integers $m^a_l$ and $n^a_l$ are specified by an integer $q^a_l$ as
\beq
m^a_l = -s_l + k_l q^a_l \ ,~~~
n^a_l = N_l - 2r_l q^a_l \ ,
\label{rel_mn_1q_6d}
\eeq
for each $a$ and $l$.
The integers $N_l$, $r_l$, $s_l$, and $k_l$ for each $l$ specify the original torus 
of the Morita equivalence
for each $T^2$. 
Equations (\ref{rel_mn_1q_6d}) can be inverted as\footnote{
In the notations of $p$ and $q$ in refs.~\cite{Ambjorn:1999ts}-\cite{Aoki:2008ik},
the present case corresponds to $\tilde{p}=1$, $\tilde{q}=q^a_l$, and $p_0=p^a$,
and thus $p = p^a$ and $q= p^a q^a_l$.
Since the configurations (\ref{conf_V6}) depend only on
the dual tori,
the original torus seems virtual.
It is introduced just for treating all the dual tori equally. 
One could regard one of the dual tori as an original torus.
}
\beq
1=2r_l m^a_l + k_l n^a_l \ ,~~~
q^a_l = N_l m^a_l + s_l n^a_l \ .
\label{rel_1q_mn_6d}
\eeq
In fact, the configurations (\ref{conf_V6}),
when inserted into (\ref{relcalVV1}), are classical solutions for 
the action (\ref{TEK-action}) (see, for instance, ref.~\cite{Griguolo:2003kq}).
Their semiclassical analyses will be given in section \ref{sec:probpheno}.

The fermionic matrix $\Psi$ is similarly decomposed into blocks as
\beq
\Psi=
\begin{pmatrix}
\varphi ^{11} \otimes \psi^{11}  & \cdots & \varphi ^{1h} \otimes  \psi^{1h} \cr
\vdots  & \ddots & \vdots \cr
\varphi ^{h1} \otimes \psi^{h1} & \cdots & \varphi ^{hh} \otimes \psi^{hh}
\end{pmatrix} \ ,
\label{psi_block_decompose}
\eeq
where $\varphi^{ab}$ and  $\psi^{ab}$ represent spinor fields on $M^4$
and $T^6$, respectively.
Each block $\varphi^{ab} \otimes \psi^{ab}$ is in a bifundamental representation
$(p^a,\bar{p^b})$ under the gauge group ${\rm U}(p^a) \times {\rm U}(p^b)$.
It turns out \cite{Aoki:2010gv}
that $\psi^{ab}$ has a topological charge,
i.e., a magnetic flux, on $T^6$ as 
\beq
p^a p^b \prod_{l=1}^3 (n^b_l m^a_l - m^b_l n^a_l)=
p^a p^b \prod_{l=1}^3 (q^a_l-q^b_l) = 
p^a p^b \prod_{l=1}^3 \left(-\frac{1}{2r}(n^a_l-n^b_l)\right) \ .
\label{indexab}
\eeq
Note that
the magnetic fluxes (\ref{indexab}) are specified by the 
configurations (\ref{conf_V6}),
not by the twists ${\cal Z}_{MN}$ in the action (\ref{TEK-action}).
In fact, by defining an overlap-Dirac operator,
which satisfies a Ginsparg-Wilson relation and an index theorem\footnote{
These techniques were developed in the lattice gauge theories \cite{GinspargWilson}-\cite{Luscher}
and applied to MM and NC geometries \cite{balagovi}-\cite{AIN2}.},
the Dirac index, i.e., 
the difference between the numbers of chiral zero modes,
was shown to take the corresponding values\footnote{
The same results were obtained in the fuzzy spheres
\cite{Aoki:2010hx},\cite{AIN3}-\cite{Aoki:2009cv}.}.  
In the present paper, we do not specify the forms of the Dirac operator,
and just assume that 
the correct number of chiral zero modes arises
in the large-$N$ limit or in the low-energy effective theory. 

Although we work with the toroidal compactifications in this paper,
our phenomenological analyses in the following sections
can be performed similarly
with any compactifications with nontrivial topologies.

\section{Matrix configurations for some phenomenological models without the Higgsino fields}
\label{sec:confpheno}
\setcounter{equation}{0}

We now study matrix configurations that provide 
the matter content of some phenomenological models.

As we saw in the previous section, in the present formulation,
one can realize models with a gauge group 
${\rm U}(p^1) \times {\rm U}(p^2) \times \cdots \times {\rm U}(p^h)$
and bifundamental matter fields.
Unfortunately, as we showed in ref.~\cite{Aoki:2012ei},
in the model with the gauge group 
${\rm U}(3) \times {\rm U}(2) \times {\rm U}(1)^2 
\simeq {\rm SU}(3) \times {\rm SU}(2) \times {\rm U}(1)^4 \subset {\rm U}(7)$,
either the right-handed singlet neutrino or the B-L gauge field cannot be 
included in the model.
Neither do we have any solution for the fluxes that provide the SM fermions
with three generations, 
in any models whose gauge group is a subgroup of U(7)
or smaller groups.

We then consider models whose gauge group is a subgroup of U(8).
All the subgroups of U(8), which yield the SM fermions 
with three generations, are 
${\rm U}(4) \times {\rm U}_L(2) \times {\rm U}_R(2)$,
${\rm U}_c(3) \times {\rm U}(1) \times {\rm U}_L(2) \times {\rm U}_R(2)$,
${\rm U}(4) \times {\rm U}_L(2) \times {\rm U}(1)^2$,
and ${\rm U}_c(3) \times {\rm U}(1) \times {\rm U}_L(2) \times {\rm U}(1)^2$. 
We will examine all of them.
Since the 
${\rm U}_c(3) \times {\rm U}(1) \times {\rm U}_L(2) \times {\rm U}(1)^2$
model turns out to have no solution,
we further study a model with 
${\rm U}_c(3) \times {\rm U}_L(2) \times {\rm U}(1)^4 \subset {\rm U}(9)$,
in order to find a model where the extra gauge group is Abelian.

As for the Higgs field,
the bosonic matrices in the extra dimensions $V_i$,
when considering fluctuations around the background (\ref{conf_V6}),
give the gauge fields in the extra dimensions,
and hence scalar fields on our spacetime.
When the matrices are decomposed into blocks,
some of the block elements have the same representation 
under the gauge group as the Higgs field.
However, the off-diagonal blocks give massive fields in general.
A more serious problem is that, 
as mentioned in the introduction,
even if scalar fields are massless at the tree level, 
it is difficult to keep them massless against quantum corrections,
which is well known as the naturalness or the hierarchy problem.
In the gauge-Higgs unifications \cite{Manton:1979kb}-\cite{Lim:2006bx}, 
higher-dimensional gauge symmetries
protect the scalar mass from the quadratic divergences of the cutoff order,
but it still can receive quantum corrections of the order of the Kaluza-Klein scale.
See also ref.~\cite{Aoki:2012xs} for an argument of the quadratic divergences.

In this section, we will find matrix configurations that provide the gauge fields and
the SM fermions,
assuming that the above-mentioned Higgs candidates are kept massless by some mechanisms.

\subsection{${\rm U}(4) \times {\rm U}_L(2) \times {\rm U}_R(2)$ model}
\label{sec:model422}

We first consider the case with
the number of blocks being three, i.e., $h=3$,
and the integers $p^a$ taken to be $4,2,2$ for $a=1, \ldots , h$.
The gauge group is then ${\rm U}(4) \times {\rm U}(2) \times {\rm U}(2)$.

The fermionic species are embedded in the fermionic matrix $\psi$ as
\beq
\psi = \left(
\begin{array}{c|c|c}
o&\begin{array}{c}q \\ l \end{array}&
\begin{array}{c}u~d \\ \nu~e \end{array} \\ \hline
&o& o \\ \hline
&&o 
\end{array} 
\right) \ ,
\label{fermionembed422}
\eeq
where
$q$ denotes the quark doublets,
$l$ the lepton doublets,
$u$ and $d$ the quark singlets,
and $\nu$ and $e$ the lepton singlets.
The entries denoted as $o$ give no massless fermions
since, as we will see below,
they are set to have a vanishing index.
The lower triangle part can be
obtained from the upper part by the charge conjugation transformation.

The fields $q$ and $l$ are in the $(4,\bar{2},1)$ representation 
under the gauge group ${\rm U}(4) \times {\rm U}_L(2) \times {\rm U}_R(2)$.
The fields $u$, $d$, $\nu$, and $e$ are in $(4,1,\bar{2})$.
As we will see in section \ref{sec:model31211},
the fermionic fields have the correct representation under the SM gauge group
${\rm SU}_c(3) \times {\rm SU}_L(2) \times {\rm U}_Y(1)$,
which is a subgroup of ${\rm U}(4) \times {\rm U}_L(2) \times {\rm U}_R(2)$.
The color group ${\rm SU}_c(3)$ and the lepton number U(1) are unified to
${\rm SU}(4)$, which is reminiscent of the Pati-Salam model \cite{Pati:1974yy}.

We now determine the integers $q^a_l$ specifying the magnetic fluxes.
From (\ref{indexab}), only the differences $q^a_l-q^b_l$ are relevant to the 
topology for the block $\psi^{ab}$.
We thus define
\beqa
q^{ab}_l &=& q^a_l-q^b_l \ ,
\label{defqlab}\\
q^{ab} &=& \prod_{l=1}^3 q^{ab}_l \ .
\label{defqab}
\eeqa
In order for (\ref{fermionembed422}) to have the correct generation number,
the integers $q^{ab}$ must have the values 
\beq
q^{ab}=
\begin{pmatrix}
0&-3&3\cr
&0&0\cr
&&0
\end{pmatrix} \ .
\label{q_ab_422}
\eeq
The lower triangle part is obtained from the upper part
by the relation $q^{ab}= - q^{ba}$.
The block component with a vanishing index 
gives no chiral zero modes, and thus no massless fermions on our spacetime.
Even when there are right-handed and left-handed zero modes in a pair,
they will obtain a mass through quantum corrections,
though they are massless at the tree level.

The fluxes in each $T^2$, $q^{ab}_l$, are obtained by solving
eq.~(\ref{defqab}) with the $T^6$ fluxes (\ref{q_ab_422}).
Here, we note that eq.~(\ref{defqab}) is invariant under
the permutations among $q^{ab}_1$, $q^{ab}_2$, and $q^{ab}_3$,
and also under the sign flips:
$q^{ab}_1 \to -q^{ab}_1$, $q^{ab}_2 \to -q^{ab}_2$, $q^{ab}_3 \to q^{ab}_3$;
$q^{ab}_1 \to -q^{ab}_1$, $q^{ab}_2 \to q^{ab}_2$, $q^{ab}_3 \to -q^{ab}_3$;
$q^{ab}_1 \to q^{ab}_1$, $q^{ab}_2 \to -q^{ab}_2$, $q^{ab}_3 \to -q^{ab}_3$.
By using these symmetries, we can fix the order of 
$q^{ab}_1$, $q^{ab}_2$, and $q^{ab}_3$,
and their overall signs.
Under this constraint, there are four solutions for $q^{ab}_l$.
We list them in Table \ref{table:q_ab_l_422}.
In order to save space,
we have omitted the diagonal elements that always vanish.
We will call these matrices $\hat{q}^{ab}_l$.

\begin{table}
\begin{center}
\begin{tabular}{c|c|c}
$\hat{q}_1^{ab}$ & $\hat{q}_2^{ab}$ & $\hat{q}_3^{ab}$ \\ \hline\hline
$\begin{pmatrix}
-1&1\cr
&2\cr
\end{pmatrix}$
&
$\begin{pmatrix}
1&1\cr
&0\cr
\end{pmatrix}$
&
$\begin{pmatrix}
3&3\cr
&0\cr
\end{pmatrix}$ \\ \hline
$\begin{pmatrix}
-1&-1\cr
&0\cr
\end{pmatrix}$
&
$\begin{pmatrix}
1&1\cr
&0\cr
\end{pmatrix}$
&
$\begin{pmatrix}
3&-3\cr
&-6\cr
\end{pmatrix}$ \\ \hline
$\begin{pmatrix}
-1&-1\cr
&0\cr
\end{pmatrix}$
&
$\begin{pmatrix}
1&-3\cr
&-4\cr
\end{pmatrix}$
&
$\begin{pmatrix}
3&1\cr
&-2\cr
\end{pmatrix}$ \\ \hline
$\begin{pmatrix}
-1&-1\cr
&0\cr
\end{pmatrix}$
&
$\begin{pmatrix}
1&3\cr
&2\cr
\end{pmatrix}$
&
$\begin{pmatrix}
3&-1\cr
&-4\cr
\end{pmatrix}$ 
\end{tabular} 
\caption{Fluxes in each $T^2$ in the ${\rm U}(4) \times {\rm U}_L(2) \times {\rm U}_R(2)$ model,
i.e., all the
solutions for eq.~(\ref{defqab}) with the $T^6$ fluxes (\ref{q_ab_422}).
The diagonal elements are omitted.}
\label{table:q_ab_l_422}
\end{center}
\end{table}

\subsection{${\rm U}_c(3) \times {\rm U}(1) \times {\rm U}_L(2) \times {\rm U}_R(2)$ model}
\label{sec:model3122}

We then consider the case where the gauge group ${\rm U}(4)$ in the previous subsection
is broken down to ${\rm U}_c(3) \times {\rm U}(1)$ by the fluxes, i.e.,
the case with $h=4$ and $p^a=(3,1,2,2)$.

The fermionic species are embedded in the fermionic matrix $\psi$ as
\beq
\psi = \left(
\begin{array}{c|c|c|c}
o&o&q&u~d \\ \hline
&o&l&\nu~e \\ \hline
&&o& o \\ \hline
&&&o 
\end{array} 
\right) \ .
\label{fermionembed3122}
\eeq
The fluxes in $T^6$ must have the values 
\beq
q^{ab}=
\begin{pmatrix}
0&0&-3&3\cr
&0&-3&3\cr
&&0&0\cr
&&&0
\end{pmatrix} \ .
\label{q_ab_3122}
\eeq
The fluxes in each $T^2$, $q^{ab}_l$, are obtained by solving
eq.~(\ref{defqab}) with the $T^6$ fluxes (\ref{q_ab_3122}).
They can be obtained by doubling the first row of the matrices in Table \ref{table:q_ab_l_422}.
However, if $q^{12}_l=0$ for all $l$,
which is equivalent to $q^1_l=q^2_l$ for all $l$,
the first diagonal block and the second one 
in the bosonic configuration (\ref{conf_V6}) become identical,
and the gauge group is enhanced from ${\rm U}(3) \times {\rm U}(1)$ to ${\rm U}(4)$,
which brings us back to the case in the previous subsection.
Hence, we must find a solution that has both $q^{12}_l=0$ and $q^{12}_l \ne 0$, 
depending on $l$.
There are four solutions,
which we list in Table \ref{table:q_ab_l_3122}.

\begin{table}
\begin{center}
\begin{tabular}{c|c|c}
$\hat{q}_1^{ab}$ & $\hat{q}_2^{ab}$ & $\hat{q}_3^{ab}$ \\ \hline\hline
$\begin{pmatrix}
-2&-1&-1\cr
&1&1\cr
&&0
\end{pmatrix}$
&
$\begin{pmatrix}
-2&-1&-1\cr
&1&1\cr
&&0
\end{pmatrix}$
&
$\begin{pmatrix}
0&-3&3\cr
&-3&3\cr
&&6
\end{pmatrix}$
\\ \hline
$\begin{pmatrix}
0&-1&1\cr
&-1&1\cr
&&2
\end{pmatrix}$
&
$\begin{pmatrix}
2&1&1\cr
&-1&-1\cr
&&0
\end{pmatrix}$
&
$\begin{pmatrix}
6&3&3\cr
&-3&-3\cr
&&0
\end{pmatrix}$
\\ \hline
$\begin{pmatrix}
0&-1&1\cr
&-1&1\cr
&&2
\end{pmatrix}$
&
$\begin{pmatrix}
4&3&3\cr
&-1&-1\cr
&&0
\end{pmatrix}$
&
$\begin{pmatrix}
4&1&1\cr
&-3&-3\cr
&&0
\end{pmatrix}$
\\ \hline
$\begin{pmatrix}
0&-1&1\cr
&-1&1\cr
&&2
\end{pmatrix}$
&
$\begin{pmatrix}
2&3&3\cr
&1&1\cr
&&0
\end{pmatrix}$
&
$\begin{pmatrix}
-2&1&1\cr
&3&3\cr
&&0
\end{pmatrix}$
\end{tabular} 
\caption{Fluxes in each $T^2$ in the 
${\rm U}_c(3) \times {\rm U}(1) \times {\rm U}_L(2) \times {\rm U}_R(2)$ model,
i.e., all the
solutions for eq.~(\ref{defqab}) 
with the $T^6$ fluxes (\ref{q_ab_3122}).
The diagonal elements are omitted.}
\label{table:q_ab_l_3122}
\end{center}
\end{table}

\subsection{${\rm U}(4) \times {\rm U}_L(2) \times {\rm U}(1)^2$ model}
\label{sec:model4211}

We next consider the case where the gauge group ${\rm U}_R(2)$ 
is broken down to ${\rm U}(1) \times {\rm U}(1)$ by the fluxes, i.e.,
the case with $h=4$ and $p^a=(4,2,1,1)$.

The fermionic species are embedded in the fermionic matrix $\psi$ as
\beq
\psi =
\left(
\begin{array}{c|c|c|c}
o&\begin{array}{c}q \\ l \end{array}&
\begin{array}{c}u \\ \nu \end{array}&
\begin{array}{c}d \\ e \end{array} \\ \hline
&o&o&o \\ \hline
&&o&o \\ \hline
&&&o 
\end{array} 
\right) \ .
\label{fermionembed4211}
\eeq
The fluxes in $T^6$ must have the values 
\beq
q^{ab}=
\begin{pmatrix}
0&-3&3&3\cr
&0&0&0\cr
&&0&0\cr
&&&0
\end{pmatrix} \ .
\label{q_ab_4211}
\eeq
The fluxes in each $T^2$, $q^{ab}_l$, are obtained by solving
eq.~(\ref{defqab}) with the $T^6$ fluxes (\ref{q_ab_4211}).
They can be obtained by suitably doubling the last column of the matrices in
Table~\ref{table:q_ab_l_422}.
Under the constraints mentioned in the previous subsections,
there are nine solutions, where we have also fixed the exchange between the last two columns.
We list them in Table \ref{table:q_ab_l_4211}.

\begin{table}
\begin{center}
\begin{tabular}{c|c|c}
$\hat{q}_1^{ab}$ & $\hat{q}_2^{ab}$ & $\hat{q}_3^{ab}$ \\ \hline\hline
$\begin{pmatrix}
-1&1&-1\cr
&2&0\cr
&&-2
\end{pmatrix}$
&
$\begin{pmatrix}
1&1&-1\cr
&0&-2\cr
&&-2
\end{pmatrix}$
&
$\begin{pmatrix}
3&3&3\cr
&0&0\cr
&&0
\end{pmatrix}$
\\ \hline
$\begin{pmatrix}
-1&1&-1\cr
&2&0\cr
&&-2
\end{pmatrix}$
&
$\begin{pmatrix}
1&1&1\cr
&0&0\cr
&&0
\end{pmatrix}$
&
$\begin{pmatrix}
3&3&-3\cr
&0&-6\cr
&&-6
\end{pmatrix}$
\\ \hline
$\begin{pmatrix}
-1&1&3\cr
&2&4\cr
&&2
\end{pmatrix}$
&
$\begin{pmatrix}
1&1&1\cr
&0&0\cr
&&0
\end{pmatrix}$
&
$\begin{pmatrix}
3&3&1\cr
&0&-2\cr
&&-2
\end{pmatrix}$
\\ \hline
$\begin{pmatrix}
-1&1&-3\cr
&2&-2\cr
&&-4
\end{pmatrix}$
&
$\begin{pmatrix}
1&1&1\cr
&0&0\cr
&&0
\end{pmatrix}$
&
$\begin{pmatrix}
3&3&-1\cr
&0&-4\cr
&&-4
\end{pmatrix}$
\\ \hline
$\begin{pmatrix}
-1&-1&3\cr
&0&4\cr
&&4
\end{pmatrix}$
&
$\begin{pmatrix}
1&1&1\cr
&0&0\cr
&&0
\end{pmatrix}$
&
$\begin{pmatrix}
3&-3&1\cr
&-6&-2\cr
&&4
\end{pmatrix}$
\\ \hline
$\begin{pmatrix}
-1&-1&-3\cr
&0&-2\cr
&&-2
\end{pmatrix}$
&
$\begin{pmatrix}
1&1&1\cr
&0&0\cr
&&0
\end{pmatrix}$
&
$\begin{pmatrix}
3&-3&-1\cr
&-6&-4\cr
&&2
\end{pmatrix}$
\\ \hline
$\begin{pmatrix}
-1&-1&3\cr
&0&4\cr
&&4
\end{pmatrix}$
&
$\begin{pmatrix}
1&-3&1\cr
&-4&0\cr
&&4
\end{pmatrix}$
&
$\begin{pmatrix}
3&1&1\cr
&-2&-2\cr
&&0
\end{pmatrix}$
\\ \hline
$\begin{pmatrix}
-1&-1&-1\cr
&0&0\cr
&&0
\end{pmatrix}$
&
$\begin{pmatrix}
1&-3&3\cr
&-4&2\cr
&&6
\end{pmatrix}$
&
$\begin{pmatrix}
3&1&-1\cr
&-2&-4\cr
&&-2
\end{pmatrix}$
\\ \hline
$\begin{pmatrix}
-1&-1&-3\cr
&0&-2\cr
&&-2
\end{pmatrix}$
&
$\begin{pmatrix}
1&3&1\cr
&2&0\cr
&&-2
\end{pmatrix}$
&
$\begin{pmatrix}
3&-1&-1\cr
&-4&-4\cr
&&0
\end{pmatrix}$
\end{tabular} 
\caption{Fluxes in each $T^2$ in the 
${\rm U}(4) \times {\rm U}_L(2) \times {\rm U}(1) \times {\rm U}(1)$ model,
i.e., all the
solutions for eq.~(\ref{defqab}) with the $T^6$ fluxes (\ref{q_ab_4211}).
The diagonal elements are omitted.}
\label{table:q_ab_l_4211}
\end{center}
\end{table}

\subsection{${\rm U}_c(3) \times {\rm U}(1) \times {\rm U}_L(2) \times {\rm U}(1)^2$ model} 
\label{sec:model31211}

We now consider the case where both 
${\rm U}(4) \to {\rm U}_c(3) \times {\rm U}(1)$ and ${\rm U}_R(2) \to {\rm U}(1)^2$ 
take place by the fluxes, i.e.,
the case with $h=5$ and $p^a=(3,1,2,1,1)$.

The fermionic species are embedded in the fermionic matrix $\psi$ as
\beq
\psi = \left(
\begin{array}{c|c|c|c|c}
o&o&q&u&d \\ \hline
&o&l&\nu&e \\ \hline
&&o& o&o \\ \hline
&&&o &o \\ \hline
&&&&o
\end{array} 
\right) \ .
\label{fermionembed31211}
\eeq
The fluxes in $T^6$ must have the values 
\beq
q^{ab}=
\begin{pmatrix}
0&0&-3&3&3\cr
&0&-3&3&3\cr
&&0&0&0\cr
&&&0&0\cr
&&&&0
\end{pmatrix} \ .
\label{q_ab_31211}
\eeq
The fluxes in each $T^2$, $q^{ab}_l$, are obtained by solving
eq.~(\ref{defqab}) with the $T^6$ fluxes (\ref{q_ab_31211}).
They could be obtained by suitably doubling the last column of the matrices in 
Table~\ref{table:q_ab_l_3122},
or by suitably doubling the first row of the matrices in
Table~\ref{table:q_ab_l_4211}.
Unfortunately, however, there is no solution.

We then generalize the fermion embedding (\ref{fermionembed31211}) to
\beq
\psi =
\begin{pmatrix}
o&u'&q&u&d \cr 
&o&l&\nu(\bar{\nu})&e \cr
&&o&\bar{l'}&o \cr
&&&o &e' \cr 
&&&&o
\end{pmatrix}  \ ,
\label{fermionembed31211I} \\
\eeq
\beq
\psi =
\begin{pmatrix}
o&u&q&d_1&d_2 \cr 
&o&l&\e_1&e_2 \cr
&&o&o&o \cr
&&&o &\nu(\bar{\nu}) \cr 
&&&&o
\end{pmatrix}  \ ,
\label{fermionembed31211II}
\eeq
though they cannot be applied to the models with
larger gauge groups in the previous subsections.
Here, we have omitted the lines that separate the matrix elements
since they are no longer necessary.

In fact, (\ref{fermionembed31211I}) and (\ref{fermionembed31211II})
are the most general embeddings where all the elements have the correct representation
under the SM gauge group ${\rm SU}_c(3) \times {\rm SU}_L(2) \times {\rm U}_Y(1)$:
One can easily see that
they have the correct 
representation under ${\rm SU}_c(3) \times {\rm SU}_L(2)$.
Since we have five ${\rm U}(1)$ gauge groups coming from each diagonal block,
we consider their linear combinations
\beq
\sum_{i=1}^5 x^i Q^i \ ,
\label{lcxQ}
\eeq
where $Q^i$ is the ${\rm U}(1)$ charge from the $i$th block.
From the hypercharge assignment for the fermion species,
the only possible embeddings turn out to be
those in (\ref{fermionembed31211I}) and (\ref{fermionembed31211II}).

For the (\ref{fermionembed31211I}) case,
we can consider the
hypercharge $Y$, baryon number $B$, lepton number $L'$,
left-handed charge $Q_L$, and right-handed charge $Q_R'$.
Their charges for $q$, $u$, $u'$, $d$, $l$, $l'$, $\nu (\bar\nu)$, $e$, and $e'$, 
and the corresponding values for $x^i$ are given 
in Table \ref{table:U(1)charges31211I}.
The charges $L'$ and $Q'_R$ have the proper interpretation when
$u'$, $l'$, and $e'$ disappear, and $\nu$, not $\bar\nu$, is chosen 
in Table~\ref{table:U(1)charges31211I}, and thus in (\ref{fermionembed31211I}).
This case reduces to the original embedding (\ref{fermionembed31211}).
This also ensures that the fermion species 
in (\ref{fermionembed422}), (\ref{fermionembed3122}), and (\ref{fermionembed4211})
in the previous subsections have the correct representation under
the SM gauge group and the extra U(1) gauge groups,
which are subgroups of the gauge group in the corresponding model.

\begin{table}
\begin{center}
\begin{tabular}{c||c|c|c|c|c|c|c|c|c}
&$q$&$u$&$u'$&$d$&$l$&$l'$&$\nu(\bar\nu)$&$e$&$e'$ \\ \hline\hline
$Y$&1/6&2/3&2/3&-1/3&-1/2&-1/2&0&-1&-1 \\ \hline
$B$&1/3&1/3&1/3&1/3&0&0&0&0&0 \\ \hline
$L'$&0&0&-1&0&1&0&1&1&0 \\ \hline
$Q_L$&1&0&0&0&1&1&0&0&0 \\ \hline
$Q'_R$&0&1&0&1&0&-1&1&1&0 
\end{tabular} \\
\begin{tabular}{c||c|c|c|c|c}
&$x^1$&$x^2$&$x^3$&$x^4$&$x^5$ \\ \hline\hline
$Y$&1/6&-1/2&0&-1/2&1/2 \\ \hline
$B$&1/3&0&0&0&0 \\ \hline
$L'$&0&1&0&0&0 \\ \hline
$Q_L$&0&0&-1&0&0 \\ \hline
$Q'_R$&0&0&0&-1&-1 
\end{tabular} 
\caption{${\rm U}(1)$ charges for $q$, $u$, $u'$, $d$, $l$, $l'$, $\nu (\bar\nu)$, $e$, and $e'$
in (\ref{fermionembed31211I}), and the corresponding values for the coefficients $x^i$
in (\ref{lcxQ}).}
\label{table:U(1)charges31211I}
\end{center}
\end{table}

For the (\ref{fermionembed31211II}) case,
we consider the
hypercharge $Y$, baryon number $B$, left-handed charge $Q_L$, 
and two other charges $Q'_1$ and $Q'_2$.
Their charges for $q$, $u$, $d_1$, $d_2$, $l$, $\nu (\bar\nu)$, $e_1$, and $e_2$, 
and the corresponding values for $x^i$ are given 
in Table \ref{table:U(1)charges31211II}.
The charges $Q'_1$ and $Q'_2$ have no proper interpretations.
We also note that, in the cases (\ref{fermionembed31211I}) and (\ref{fermionembed31211II}),
a linear combination of these five ${\rm U}(1)$ charges
gives an overall ${\rm U}(1)$ 
and does not couple to the matter.
Only four ${\rm U}(1)$ charges couple to the matter.

\begin{table}
\begin{center}
\begin{tabular}{c||c|c|c|c|c|c|c|c}
&$q$&$u$&$d_1$&$d_2$&$l$&$\nu(\bar\nu)$&$e_1$&$e_2$ \\ \hline\hline
$Y$&1/6&2/3&-1/3&-1/3&-1/2&0&-1&-1 \\ \hline
$B$&1/3&1/3&1/3&1/3&0&0&0&0 \\ \hline
$Q_L$&1&0&0&0&1&0&0&0 \\ \hline
$Q'_1$&0&0&1&0&0&-1&1&0 \\ \hline
$Q'_2$&0&0&0&1&0&1&0&1
\end{tabular} \\
\begin{tabular}{c||c|c|c|c|c}
&$x^1$&$x^2$&$x^3$&$x^4$&$x^5$ \\ \hline\hline
$Y$&1/6&-1/2&0&1/2&1/2 \\ \hline
$B$&1/3&0&0&0&0 \\ \hline
$Q_L$&0&0&-1&0&0 \\ \hline
$Q'_1$&0&0&0&-1&0 \\ \hline
$Q'_2$&0&0&0&0&-1 
\end{tabular} 
\caption{${\rm U}(1)$ charges for  
$q$, $u$, $d_1$, $d_2$, $l$, $\nu (\bar\nu)$, $e_1$, and $e_2$,
in (\ref{fermionembed31211II}), and the corresponding values for the coefficients $x^i$
in (\ref{lcxQ}).}
\label{table:U(1)charges31211II}
\end{center}
\end{table}

The fluxes in $T^6$ must have the values
\beq
q^{ab}=
\begin{pmatrix}
0&x&-3&3-x&3 \cr
&0&y-3&\pm 3&3-z\cr
&&0&y&0\cr
&&&0&z\cr
&&&&0
\end{pmatrix} 
\label{q_ab_31211I}
\eeq
for (\ref{fermionembed31211I}), and
\beq
q^{ab}=
\begin{pmatrix}
0&3&-3&x&3-x \cr
&0&-3&y&3-y\cr
&&0&0&0\cr
&&&0&\pm 3\cr
&&&&0
\end{pmatrix} 
\label{q_ab_31211II}
\eeq
for (\ref{fermionembed31211II}).
Here, $x$, $y$, and $z$ can take an integer $0$, $1$, $2$, or $3$.
Since the up-type quarks are embedded at the two places $u$ and $u'$ in 
(\ref{fermionembed31211I}), the corresponding fluxes 
can take several values $x$ and $3-x$ in (\ref{q_ab_31211I}).
The same is true for $l$ and $l'$, and so on.
The double signs of $\pm 3$ are chosen depending on whether $\nu$ or $\bar\nu$ is embedded in
(\ref{fermionembed31211I}) and (\ref{fermionembed31211II}). 
The fluxes in each $T^2$, $q^{ab}_l$, are obtained by solving
eq.~(\ref{defqab}) with the $T^6$ fluxes (\ref{q_ab_31211I}) and (\ref{q_ab_31211II}).
As we will show in Appendix~\ref{sec:flux31211},
there is no solution, either.
We thus conclude that there is no solution in
the ${\rm U}_c(3) \times {\rm U}(1) \times {\rm U}_L(2) \times {\rm U}(1)^2$ model.

\subsection{${\rm U}_c(3) \times {\rm U}_L(2) \times {\rm U}(1)^4$ model}
\label{sec:model321111}

We then consider the case with the gauge group  
${\rm U}_c(3) \times {\rm U}_L(2) \times {\rm U}(1)^4 \subset {\rm U}(9)$,
i.e., the case with $h=6$ and $p^a=(3,2,1,1,1,1)$,
in order to find a model with the SM gauge group plus extra U(1).

The fermionic species are embedded in the fermionic matrix $\psi$ as
\beq
\psi =
\begin{pmatrix}
o&q&u&u&u&d \cr 
&o&\bar{l}&\bar{l}&\bar{l}&o \cr
&&o&\nu&\nu&e \cr
&&&o &\nu&e \cr 
&&&&o&e \cr
&&&&&o
\end{pmatrix}  \ ,
\label{fermionembed321111I} 
\eeq
\beq
\psi =
\begin{pmatrix}
o&q&u&u&d&d \cr 
&o&\bar{l}&\bar{l}&o&o \cr
&&o&\nu&e&e \cr
&&&o &e&e \cr 
&&&&o&\nu \cr
&&&&&o
\end{pmatrix}  \ ,
\label{fermionembed321111II} 
\eeq
\beq
\psi =
\begin{pmatrix}
o&q&u&d&d&d \cr 
&o&\bar{l}&o&o&o \cr
&&o&e&e&e \cr
&&&o &\nu&\nu \cr 
&&&&o&\nu \cr
&&&&&o
\end{pmatrix}  \ ,
\label{fermionembed321111III}
\eeq
where $\nu$ can be either $\nu$ or $\bar{\nu}$.
They exhaust all the embeddings that have the correct representation
under the SM gauge group.

The fluxes in $T^6$, $q^{ab}$, can take several values
for (\ref{fermionembed321111I}),
(\ref{fermionembed321111II}), and
(\ref{fermionembed321111III}),
as in (\ref{q_ab_31211I}) and (\ref{q_ab_31211II}).
We can also determine the fluxes for $T^2$, $q^{ab}_l$,
by solving eq.~(\ref{defqab}).
As we will show in Appendix~\ref{sec:flux321111},
there are fifteen solutions for the case (\ref{fermionembed321111I}).
The $T^2$ fluxes are listed in Tables~\ref{table:q_ab_l_321111I1}, 
\ref{table:q_ab_l_321111I2}, and \ref{table:q_ab_l_321111I3}.
The corresponding $T^6$ fluxes are given in
(\ref{q_ab_321111I1}) to (\ref{q_ab_321111I6}).

For the case (\ref{fermionembed321111II}),
there are two solutions.
The $T^2$ fluxes are listed in 
Table~\ref{table:q_ab_l_321111II}.
The corresponding $T^6$ fluxes are
(\ref{q_ab_321111II1}) and (\ref{q_ab_321111II2}).
For the case (\ref{fermionembed321111III}), there is no solution. 

\section{Matrix configurations for the phenomenological models with the Higgsino fields}
\label{sec:conhiggs}
\setcounter{equation}{0}

In this section,
we assume situations where
the Higgs mass is protected by the supersymmetry
possessed by the IIB MM somehow.
Then, we will find matrix configurations 
that yield candidates for the Higgsino fields
as well as the gauge fields and the SM fermions.
We just find candidates for the fermionic fields.
How the supersymmetry is realized in the unitary MM and in the whole spectrum
will be studied elsewhere.

\subsection{${\rm U}(4) \times {\rm U}_L(2) \times {\rm U}_R(2)$ model}

We first consider the model
studied in section~\ref{sec:model422}.
The Higgsino fields $h_u$ and $h_d$ can be added to the fermion
embedding (\ref{fermionembed422}) as\footnote{
A similar embedding was given in refs.~\cite{Abe:2012ya,Abe:2012fj}.}
\beq
\psi = \left(
\begin{array}{c|c|c}
o&\begin{array}{c}q \\ l \end{array}&
\begin{array}{c}u~d \\ \nu~e \end{array} \\ \hline
&o& h_u~h_d \\ \hline
&&o 
\end{array} 
\right) \ .
\label{fermionHigssembed422}
\eeq
The $T^6$ fluxes must be altered from (\ref{q_ab_422}) to
\beq
q^{ab}=
\begin{pmatrix}
0&-3&3\cr
&0&q^H\cr
&&0
\end{pmatrix} \ ,
\label{q_ab_higgs422}
\eeq
with $q^{H} \ne 0$.
The gaugino fields, $o$ in (\ref{fermionHigssembed422}),
are also protected to be massless by the gauge symmetry
and the supersymmetry.
The $T^2$ fluxes are obtained by solving
eq.~(\ref{defqab}) with the $T^6$ fluxes (\ref{q_ab_higgs422}).
There are three solutions. 
They are listed in Table~\ref{table:q_ab_l_422higgs},
where we also write the values of $q^{H}$.
Unfortunately, there is no solution with $q^{H}=1$.

\begin{table}
\begin{center}
\begin{tabular}{c|c|c||c}
$\hat{q}_1^{ab}$ & $\hat{q}_2^{ab}$ & $\hat{q}_3^{ab}$ & $q^H$
\\ \hline\hline
$\begin{pmatrix}
-1&1\cr
&2\cr
\end{pmatrix}$
&
$\begin{pmatrix}
1&-1\cr
&-2\cr
\end{pmatrix}$
&
$\begin{pmatrix}
3&-3\cr
&-6\cr
\end{pmatrix}$ 
& $24$
\\ \hline
$\begin{pmatrix}
-1&1\cr
&2\cr
\end{pmatrix}$
&
$\begin{pmatrix}
1&3\cr
&2\cr
\end{pmatrix}$
&
$\begin{pmatrix}
3&1\cr
&-2\cr
\end{pmatrix}$ 
&$-8$
\\ \hline
$\begin{pmatrix}
-1&1\cr
&2\cr
\end{pmatrix}$
&
$\begin{pmatrix}
1&-3\cr
&-4\cr
\end{pmatrix}$
&
$\begin{pmatrix}
3&-1\cr
&-4\cr
\end{pmatrix}$
&$32$
\end{tabular} 
\caption{Fluxes in each $T^2$ in the 
${\rm U}(4) \times {\rm U}_L(2) \times {\rm U}_R(2)$ model
with the Higgsino candidates,
i.e., all the solutions for eq.~(\ref{defqab}) with the $T^6$ fluxes
(\ref{q_ab_higgs422}).
The diagonal elements are omitted.
The $T^6$ flux for the Higgsinos, $q^H$, is
also listed.
}
\label{table:q_ab_l_422higgs}
\end{center}
\end{table}

\subsection{${\rm U}_c(3) \times {\rm U}(1) \times {\rm U}_L(2) \times {\rm U}_R(2)$ model}

We then consider the model
studied in section~\ref{sec:model3122}.
The Higgsino fields $h_u$ and $h_d$ are added to the 
embedding (\ref{fermionembed3122}) as
\beq
\psi = \left(
\begin{array}{c|c|c|c}
o&o&q&u~d \\ \hline
&o&l&\nu~e \\ \hline
&&o& h_u~h_d \\ \hline
&&&o 
\end{array} 
\right) \ .
\label{fermionhiggsembed3122}
\eeq
The fluxes in $T^6$ have the values 
\beq
q^{ab}=
\begin{pmatrix}
0&0&-3&3\cr
&0&-3&3\cr
&&0&q^H\cr
&&&0
\end{pmatrix} \ ,
\label{q_ab_higgs3122}
\eeq
with $q^H \ne 0$.
Solutions for the $T^2$ fluxes can be obtained by suitably doubling the
first row of the matrices in Table~\ref{table:q_ab_l_422higgs}.
There are four solutions.
They are listed in Table~\ref{table:q_ab_l_3122higgs},
where we also give the values of $q^{H}$.

Because the lepton doublet $l$ and the Higgsino $\bar{h_u}$ 
in (\ref{fermionhiggsembed3122}) have the same
representation under the SM gauge group,
they can substitute for each other.
Such Higgs fields can also give rise to electroweak symmetry breaking,
though they cannot make Yukawa couplings, 
which are prohibited by the representation of the extra gauge group.
Then, the $T^6$ fluxes can have more general values
\beq
q^{ab}=
\begin{pmatrix}
0&0&-3&3\cr
&0&q^L&3\cr
&&0&q^H\cr
&&&0
\end{pmatrix} \ .
\label{q_ab_higgs3122_2}
\eeq
Solutions for the $T^2$ fluxes can be obtained by using the
the matrices in Table~\ref{table:q_ab_l_422higgs}.
There are fourteen solutions, other than those in Table~\ref{table:q_ab_l_3122higgs}.
We do not present them here, in order to save space.

\begin{table}
\begin{center}
\begin{tabular}{c|c|c||c}
$\hat{q}_1^{ab}$ & $\hat{q}_2^{ab}$ & $\hat{q}_3^{ab}$ & $q^H$
\\ \hline\hline
$\begin{pmatrix}
2&1&3\cr
&-1&1\cr
&&2\cr
\end{pmatrix}$
&
$\begin{pmatrix}
-2&-1&1\cr
&1&3\cr
&&2\cr
\end{pmatrix}$
&
$\begin{pmatrix}
0&3&1\cr
&3&1\cr
&&-2\cr
\end{pmatrix}$ 
&$-8$
\\ \hline
$\begin{pmatrix}
0&-1&1\cr
&-1&1\cr
&&2\cr
\end{pmatrix}$
&
$\begin{pmatrix}
-4&-3&-1\cr
&1&3\cr
&&2\cr
\end{pmatrix}$
&
$\begin{pmatrix}
-4&-1&-3\cr
&3&1\cr
&&-2\cr
\end{pmatrix}$ 
&$-8$
\\ \hline
$\begin{pmatrix}
-2&-3&-1\cr
&-1&1\cr
&&2\cr
\end{pmatrix}$
&
$\begin{pmatrix}
0&1&3\cr
&1&3\cr
&&2\cr
\end{pmatrix}$
&
$\begin{pmatrix}
-2&1&-1\cr
&3&1\cr
&&-2\cr
\end{pmatrix}$ 
&$-8$
\\ \hline
$\begin{pmatrix}
0&-1&1\cr
&-1&1\cr
&&2\cr
\end{pmatrix}$
&
$\begin{pmatrix}
2&3&-1\cr
&1&-3\cr
&&-4\cr
\end{pmatrix}$
&
$\begin{pmatrix}
-2&1&-3\cr
&3&-1\cr
&&-4\cr
\end{pmatrix}$
&$32$
\end{tabular} 
\caption{Fluxes in each $T^2$ in the 
${\rm U}_c(3) \times {\rm U}(1) \times {\rm U}_L(2) \times {\rm U}_R(2)$ model
with the Higgsino candidates,
i.e., all the solutions for eq.~(\ref{defqab}) with the $T^6$ fluxes
(\ref{q_ab_higgs3122}).
The diagonal elements are omitted.
The $T^6$ flux for the Higgsinos, $q^H$, is
also listed.}
\label{table:q_ab_l_3122higgs}
\end{center}
\end{table}

\subsection{${\rm U}(4) \times {\rm U}_L(2) \times {\rm U}(1)^2$ model}

We next consider the model
studied in section~\ref{sec:model4211}.
The Higgsino fields $h_u$ and $h_d$ are added to the 
embedding (\ref{fermionembed4211}) as
\beq
\left(
\begin{array}{c|c|c|c}
o&\begin{array}{c}q \\ l \end{array}&
\begin{array}{c}u \\ \nu \end{array}&
\begin{array}{c}d \\ e \end{array} \\ \hline
&o&h_u&h_d \\ \hline
&&o&o \\ \hline
&&&o 
\end{array} 
\right) \ .
\label{fermionhiggsembed4211}
\eeq
The fluxes in $T^6$ must have the values 
\beq
q^{ab}=
\begin{pmatrix}
0&-3&3&3\cr
&0&q^{H_u}&q^{H_d}\cr
&&0&0\cr
&&&0
\end{pmatrix} \ ,
\label{q_ab_higgs4211}
\eeq
with $q^{H_u}, q^{H_d} \ne 0$.
Solutions for the $T^2$ fluxes can be obtained by suitably doubling the
last column of the matrices in Table~\ref{table:q_ab_l_422higgs}.
There are five solutions.
They are listed in Table~\ref{table:q_ab_l_4211higgs},
where we also give the values of $q^{H_u}$ and  $q^{H_d}$.
Solutions with
$q^{H_u}=0$ and $q^{H_d} \ne 0$,
or $q^{H_u}\ne 0$ and $q^{H_d} = 0$, can also be obtained by suitably combining the
last column of the matrices in 
Tables~\ref{table:q_ab_l_422} and
\ref{table:q_ab_l_422higgs}.
There are eight solutions.
They are listed in Table~\ref{table:q_ab_l_4211higgs2},
where we also give the values of $q^{H_u}$ and  $q^{H_d}$.

\begin{table}
\begin{center}
\begin{tabular}{c|c|c||c}
$\hat{q}_1^{ab}$ & $\hat{q}_2^{ab}$ & $\hat{q}_3^{ab}$ & $q^{H_u}$, $q^{H_d}$
\\ \hline\hline
$\begin{pmatrix}
-1&1&1\cr
&2&2\cr
&&0\cr
\end{pmatrix}$
&
$\begin{pmatrix}
1&-1&3\cr
&-2&2\cr
&&4\cr
\end{pmatrix}$
&
$\begin{pmatrix}
3&-3&1\cr
&-6&-2\cr
&&4\cr
\end{pmatrix}$ 
& $24$, $-8$
\\ \hline
$\begin{pmatrix}
-1&1&1\cr
&2&2\cr
&&0\cr
\end{pmatrix}$
&
$\begin{pmatrix}
1&3&-3\cr
&2&-4\cr
&&-6\cr
\end{pmatrix}$
&
$\begin{pmatrix}
3&1&-1\cr
&-2&-4\cr
&&-2\cr
\end{pmatrix}$ 
&$-8$, $32$
\\ \hline
$\begin{pmatrix}
-1&1&1\cr
&2&2\cr
&&0\cr
\end{pmatrix}$
&
$\begin{pmatrix}
1&-3&-1\cr
&-4&-2\cr
&&2\cr
\end{pmatrix}$
&
$\begin{pmatrix}
3&-1&-3\cr
&-4&-6\cr
&&-2\cr
\end{pmatrix}$
&$32$, $24$
\\ \hline
$\begin{pmatrix}
-1&1&-3\cr
&2&-2\cr
&&-4\cr
\end{pmatrix}$
&
$\begin{pmatrix}
1&3&-1\cr
&2&-2\cr
&&-4\cr
\end{pmatrix}$
&
$\begin{pmatrix}
3&1&1\cr
&-2&-2\cr
&&0\cr
\end{pmatrix}$ 
&$-8$, $-8$
\\ \hline
$\begin{pmatrix}
-1&1&3\cr
&2&4\cr
&&2\cr
\end{pmatrix}$
&
$\begin{pmatrix}
1&-3&-1\cr
&-4&-2\cr
&&2\cr
\end{pmatrix}$
&
$\begin{pmatrix}
3&-1&-1\cr
&-4&-4\cr
&&0\cr
\end{pmatrix}$
&$32$, $32$
\end{tabular} 
\caption{Fluxes in each $T^2$ in the 
${\rm U}(4) \times {\rm U}_L(2) \times {\rm U}(1)^2$ model
with the Higgsino candidates,
i.e., solutions for eq.~(\ref{defqab}) with the $T^6$ fluxes
(\ref{q_ab_higgs4211})
(part 1).
The diagonal elements are omitted.
Also listed are
the $T^6$ fluxes for the Higgsinos, $q^{H_u}$ and $q^{H_d}$, 
both of which take nonzero values.}
\label{table:q_ab_l_4211higgs}
\end{center}
\end{table}

\begin{table}
\begin{center}
\begin{tabular}{c|c|c||c}
$\hat{q}_1^{ab}$ & $\hat{q}_2^{ab}$ & $\hat{q}_3^{ab}$ & $q^{H_u}$, $q^{H_d}$
\\ \hline\hline
$\begin{pmatrix}
-1&1&1\cr
&2&2\cr
&&0\cr
\end{pmatrix}$
&
$\begin{pmatrix}
1&1&-1\cr
&0&-2\cr
&&-2\cr
\end{pmatrix}$
&
$\begin{pmatrix}
3&3&-3\cr
&0&-6\cr
&&-6\cr
\end{pmatrix}$
&$0$, $24$
\\ \hline
$\begin{pmatrix}
-1&1&1\cr
&2&2\cr
&&0\cr
\end{pmatrix}$
&
$\begin{pmatrix}
1&1&3\cr
&0&2\cr
&&2\cr
\end{pmatrix}$
&
$\begin{pmatrix}
3&3&1\cr
&0&-2\cr
&&-2\cr
\end{pmatrix}$ 
& $0$, $-8$
\\ \hline
$\begin{pmatrix}
-1&1&1\cr
&2&2\cr
&&0\cr
\end{pmatrix}$
&
$\begin{pmatrix}
1&1&-3\cr
&0&-4\cr
&&-4\cr
\end{pmatrix}$
&
$\begin{pmatrix}
3&3&-1\cr
&0&-4\cr
&&-4\cr
\end{pmatrix}$ 
&$0$, $32$
\\ \hline
$\begin{pmatrix}
-1&-1&1\cr
&0&2\cr
&&2\cr
\end{pmatrix}$
&
$\begin{pmatrix}
1&1&-1\cr
&0&-2\cr
&&-2\cr
\end{pmatrix}$
&
$\begin{pmatrix}
3&-3&-3\cr
&-6&-6\cr
&&0\cr
\end{pmatrix}$
&$0$, $24$
\\ \hline
$\begin{pmatrix}
-1&-1&1\cr
&0&2\cr
&&2\cr
\end{pmatrix}$
&
$\begin{pmatrix}
1&-3&3\cr
&-4&2\cr
&&6\cr
\end{pmatrix}$
&
$\begin{pmatrix}
3&1&1\cr
&-2&-2\cr
&&0\cr
\end{pmatrix}$ 
& $0$, $-8$
\\ \hline
$\begin{pmatrix}
-1&-1&1\cr
&0&2\cr
&&2\cr
\end{pmatrix}$
&
$\begin{pmatrix}
1&-3&-3\cr
&-4&-4\cr
&&0\cr
\end{pmatrix}$
&
$\begin{pmatrix}
3&1&-1\cr
&-2&-4\cr
&&-2\cr
\end{pmatrix}$ 
&$0$, $32$
\\ \hline
$\begin{pmatrix}
-1&-1&1\cr
&0&2\cr
&&2\cr
\end{pmatrix}$
&
$\begin{pmatrix}
1&3&3\cr
&2&2\cr
&&0\cr
\end{pmatrix}$
&
$\begin{pmatrix}
3&-1&1\cr
&-4&-2\cr
&&2\cr
\end{pmatrix}$ 
& $0$, $-8$
\\ \hline
$\begin{pmatrix}
-1&-1&1\cr
&0&2\cr
&&2\cr
\end{pmatrix}$
&
$\begin{pmatrix}
1&3&-3\cr
&2&-4\cr
&&-6\cr
\end{pmatrix}$
&
$\begin{pmatrix}
3&-1&-1\cr
&-4&-4\cr
&&0\cr
\end{pmatrix}$ 
&$0$, $32$
\end{tabular} 
\caption{Fluxes in each $T^2$ in the 
${\rm U}(4) \times {\rm U}_L(2) \times {\rm U}(1)^2$ model
with the Higgsino candidates,
i.e., solutions for eq.~(\ref{defqab}) with the $T^6$ fluxes
(\ref{q_ab_higgs4211}) (part 2).
The diagonal elements are omitted.
Also listed are
the $T^6$ fluxes for the Higgsinos, $q^{H_u}$ and $q^{H_d}$, 
one of which takes nonzero values.}
\label{table:q_ab_l_4211higgs2}
\end{center}
\end{table}

\subsection{${\rm U}_c(3) \times {\rm U}(1) \times {\rm U}_L(2) \times {\rm U}(1)^2$ model} 

We now consider the model
studied in section~\ref{sec:model31211}.
The Higgsino fields $h_u$ and $h_d$ are added to the 
embedding (\ref{fermionembed31211}) as
\beq
\psi = \left(
\begin{array}{c|c|c|c|c}
o&o&q&u&d \\ \hline
&o&l&\nu&e \\ \hline
&&o& h_u&h_d \\ \hline
&&&o &o \\ \hline
&&&&o
\end{array} 
\right) \ .
\label{fermionhiggsembed31211}
\eeq
The fluxes in $T^6$ have the values 
\beq
q^{ab}=
\begin{pmatrix}
0&0&-3&3&3\cr
&0&-3&3&3\cr
&&0&q^{H_u}&q^{H_d}\cr
&&&0&0\cr
&&&&0
\end{pmatrix} \ ,
\label{q_ab_higgs31211}
\eeq
with $q^{H_u}, q^{H_d} \ne 0$.
Solutions for the $T^2$ fluxes can be obtained by suitably doubling the
last column of the matrices in Table~\ref{table:q_ab_l_3122higgs},
or by suitably doubling the
first row of the matrices in Table~\ref{table:q_ab_l_4211higgs}.
There is one solution.
It is listed in the first row of Table~\ref{table:q_ab_l_31211higgs},
where we also give the values of $q^{H_u}$ and  $q^{H_d}$.
Solutions with $q^{H_u}=0$ and $q^{H_d}\ne 0$,
or $q^{H_u}\ne 0$ and $q^{H_d}= 0$, can also be obtained by
suitably combining the last column of the matrices in Tables~\ref{table:q_ab_l_3122}
and \ref{table:q_ab_l_3122higgs},
or by suitably doubling the
first row of the matrices in Table~\ref{table:q_ab_l_4211higgs2}.
There are two solutions,
which are listed in the second and third
rows of Table~\ref{table:q_ab_l_31211higgs}.

As in the ${\rm U}_c(3) \times {\rm U}(1) \times {\rm U}_L(2) \times {\rm U}_R(2)$ model,
since $l$ and $\bar{h_u}$ have the same representation under the SM gauge group,
the $T^6$ fluxes can have more general values
\beq
q^{ab}=
\begin{pmatrix}
0&0&-3&3&3\cr
&0&q^L&\pm 3&3\cr
&&0&q^{H_u}&q^{H_d}\cr
&&&0&0\cr
&&&&0
\end{pmatrix} \ .
\label{q_ab_higgs31211_2}
\eeq
Solutions for the $T^2$ fluxes can be obtained by using 
the matrices in Tables~\ref{table:q_ab_l_4211higgs}
and \ref{table:q_ab_l_4211higgs2}.
There are three solutions other than those in Table~\ref{table:q_ab_l_31211higgs}.
We list them in Table~\ref{table:q_ab_l_31211higgs2},
where we also give the values of  $q^{H_u}$, $q^{H_d}$,
and $q^L$.
As for the double sign of $\pm3$ in (\ref{q_ab_higgs31211_2}),
all of the solutions take $+3$.

\begin{table}
\begin{center}
\begin{tabular}{c|c|c||c}
$\hat{q}_1^{ab}$ & $\hat{q}_2^{ab}$ & $\hat{q}_3^{ab}$ & $q^{H_u}$, $q^{H_d}$
\\ \hline\hline
$\begin{pmatrix}
2&1&3&-1\cr
&-1&1&-3\cr
&&2&-2\cr
&&&-4\cr
\end{pmatrix}$
&
$\begin{pmatrix}
-2&-1&1&-3\cr
&1&3&-1\cr
&&2&-2\cr
&&&-4\cr
\end{pmatrix}$
&
$\begin{pmatrix}
0&3&1&1\cr
&3&1&1\cr
&&-2&-2\cr
&&&0\cr
\end{pmatrix}$ 
&$-8$, $-8$
\\ \hline \hline
$\begin{pmatrix}
0&-1&1&1\cr
&-1&1&1\cr
&&2&2\cr
&&&0\cr
\end{pmatrix}$
&
$\begin{pmatrix}
4&3&3&1\cr
&-1&-1&-3\cr
&&0&-2\cr
&&&-2\cr
\end{pmatrix}$
&
$\begin{pmatrix}
4&1&1&3\cr
&-3&-3&-1\cr
&&0&2\cr
&&&2\cr
\end{pmatrix}$ 
&$0$, $-8$
\\ \hline
$\begin{pmatrix}
0&-1&1&1\cr
&-1&1&1\cr
&&2&2\cr
&&&0\cr
\end{pmatrix}$
&
$\begin{pmatrix}
2&3&3&-1\cr
&1&1&-3\cr
&&0&-4\cr
&&&-4\cr
\end{pmatrix}$
&
$\begin{pmatrix}
-2&1&1&-3\cr
&3&3&-1\cr
&&0&-4\cr
&&&-4\cr
\end{pmatrix}$
&$0$, $32$
\end{tabular} 
\caption{Fluxes in each $T^2$ in the 
${\rm U}_c(3) \times {\rm U}(1) \times {\rm U}_L(2) \times {\rm U}(1)^2$ model
with the Higgsino candidates, i.e.,
all the solutions for eq.~(\ref{defqab}) with the $T^6$ fluxes
(\ref{q_ab_higgs31211}).
The diagonal elements are omitted.
The $T^6$ fluxes for the Higgsinos, $q^{H_u}$ and $q^{H_u}$, are
also listed.}
\label{table:q_ab_l_31211higgs}
\end{center}
\end{table}

\begin{table}
\begin{center}
\begin{tabular}{c|c|c||c||c}
$\hat{q}_1^{ab}$ & $\hat{q}_2^{ab}$ & $\hat{q}_3^{ab}$ & $q^{H_u}$, $q^{H_d}$ & $q^L$
\\ \hline\hline
$\begin{pmatrix}
0&-1&1&1\cr
&-1&1&1\cr
&&2&2\cr
&&&0\cr
\end{pmatrix}$
&
$\begin{pmatrix}
0&1&-1&3\cr
&1&-3&1\cr
&&-2&2\cr
&&&4\cr
\end{pmatrix}$
&
$\begin{pmatrix}
-2&3&-3&1\cr
&5&-1&3\cr
&&-6&-2\cr
&&&4\cr
\end{pmatrix}$ 
&$24$, $-8$ & $-5$
\\ \hline
$\begin{pmatrix}
0&-1&1&1\cr
&-1&1&1\cr
&&2&2\cr
&&&0\cr
\end{pmatrix}$
&
$\begin{pmatrix}
-4&1&-3&-1\cr
&5&1&3\cr
&&-4&-2\cr
&&&2\cr
\end{pmatrix}$
&
$\begin{pmatrix}
-4&3&-1&-3\cr
&7&3&1\cr
&&-4&-6\cr
&&&-2\cr
\end{pmatrix}$
&$32$, $24$ & $-35$
\\\hline
$\begin{pmatrix}
4&-1&1&3\cr
&-5&-3&-1\cr
&&2&4\cr
&&&2\cr
\end{pmatrix}$
&
$\begin{pmatrix}
-4&1&-3&-1\cr
&5&1&3\cr
&&-4&-2\cr
&&&2\cr
\end{pmatrix}$
&
$\begin{pmatrix}
0&3&-1&-1\cr
&3&-1&-1\cr
&&-4&-4\cr
&&&0\cr
\end{pmatrix}$
&$32$, $32$ & $-75$
\end{tabular} 
\caption{Fluxes in each $T^2$ in the 
${\rm U}_c(3) \times {\rm U}(1) \times {\rm U}_L(2) \times {\rm U}(1)^2$ model
with the Higgsino candidates, i.e.,
all the solutions for eq.~(\ref{defqab}) with the $T^6$ fluxes
(\ref{q_ab_higgs31211_2}),
other than those listed in Table~\ref{table:q_ab_l_31211higgs}.
The diagonal elements are omitted.
The $T^6$ fluxes for the Higgsinos and the lepton doublet, $q^{H_u}$, $q^{H_u}$, and $q^L$, are
also listed.}
\label{table:q_ab_l_31211higgs2}
\end{center}
\end{table}

\section{Probability distribution over the phenomenological models}
\label{sec:probpheno}
\setcounter{equation}{0}

We now study the dynamics of the MM semiclassically,
and estimate the probabilities for the appearance of 
the phenomenological models obtained in the previous sections.

\subsection{Instanton actions}
\label{sec:InstantonAction}

We first discuss a setting of the parameters in the unitary MM (\ref{TEK-action}).
Regarding the twists ${\cal Z}_{MN}$, we take
\beqa
&&{\cal Z}_{01}=\exp{(2\pi i \frac{s_4}{N_4})}~,~~{\cal Z}_{23}=\exp{(2\pi i \frac{s_5}{N_5})} \ , \n
&&{\cal Z}_{45}=\exp{(2\pi i \frac{s_1}{N_1})}~,
~~{\cal Z}_{67}=\exp{(2\pi i \frac{s_2}{N_2})}~,~~{\cal Z}_{89}=\exp{(2\pi i \frac{s_3}{N_3})} \ ,
\label{twistsetting}
\eeqa
with a huge difference in $N_l$ between $l=4,5$ and $l=1,2,3$:
\beq
 N_4 \simeq N_5 \gg N_1 \simeq N_2 \simeq N_3 \ .
\label{anisotropy45123}
\eeq
The other twists are taken to be one.
The total matrix size ${\cal N}$ is taken to be
\beq
{\cal N} = k \prod_{l=1}^5 N_l \ .
\label{NN'Nk}
\eeq
Note that the matrix size (\ref{NN'Nk}) is $k$ times larger than
is usually expected from the integers that specify the twists 
(\ref{twistsetting}).
The assumptions (\ref{anisotropy45123}) and (\ref{NN'Nk}) may be justified
by studying a connection between the IIB MM (\ref{IIBMMaction})
and the unitary MM (\ref{TEK-action}).

We then consider the background (\ref{relcalVV1}),
where $V_\mu$ are ${\rm U}(N_4 N_5)$ matrices
and $V_i$ are ${\rm U}(k N_1 N_2 N_3)$ matrices.
These sizes of matrices are dynamically favored 
if one takes the twist parameters as in (\ref{twistsetting}),
as can be seen from the arguments (\ref{qla0con}) below.
The size of our spacetime and that of the extra dimensions
are given by $\epsilon N_l$ with $l=4,5$ and $l=1,2,3$, respectively,
where $\epsilon$ is a lattice spacing.
They thus have a huge anisotropy 
due to (\ref{anisotropy45123}).

We consider the topological configurations (\ref{conf_V6})
for $V_i$ in (\ref{relcalVV1}).
They are classical solutions for the unitary MM (\ref{TEK-action}).
In order to match the matrix sizes in 
the configurations (\ref{conf_V6}) and the action (\ref{TEK-action}),
\beq
\sum_{a=1}^h n^a_1 n^a_2 n^a_3 p^a = k \prod_{l=1}^3 N_l
\label{msizematching}
\eeq
is required.
Plugging (\ref{conf_V6}) into (\ref{TEK-action}), 
we obtain the classical action 
\beq
S_{b}= -2 \beta {\cal N} N_4 N_5 \sum_{l=1}^3 \sum_{a=1}^h 
n^a_1 n^a_2 n^a_3 p^a \cos{\left(2\pi\left(\frac{s_l}{N_l}+\frac{m^a_l}{n^a_l}\right)\right)} \ ,
\label{clac}
\eeq
where we have given only the first term in (\ref{TEK-action}).

Recall that the integers $n^a_l$ and $m^a_l$ are specified by an original torus with
the integers $N_l$ and $s_l$ via
(\ref{rel_mn_1q_6d}) or (\ref{rel_1q_mn_6d}).
We now consider the configurations whose integers $N_l$ and $s_l$ 
agree with the ones that specify the twists ${\cal Z}_{ij}$ in the action (\ref{TEK-action})
via (\ref{twistsetting}),
since the other configurations have much larger actions.
Then, from (\ref{rel_mn_1q_6d}) and (\ref{rel_1q_mn_6d}),
we can find the relation
\beq
\frac{s_l}{N_l}+\frac{m^a_l}{n^a_l} = \frac{q^a_l}{N_l n^a_l}
= -\frac{1}{2 r}\left(\frac{1}{N_l}-\frac{1}{n^a_l}\right) \ .
\label{relsNmnq}
\eeq
By plugging (\ref{relsNmnq}) into (\ref{clac}),
we find that
the classical action (\ref{clac}) takes the minimum value 
if and only if
\beq
q^a_l=0 \Leftrightarrow n^a_l =N_l
\label{qla0con}
\eeq
for $\forall a$ and $\forall l$. 
Then, the constraint (\ref{msizematching}) becomes
\beq
\sum_{a=1}^h p^a =k \ .
\label{blocknumberkcon}
\eeq
Therefore, if we choose the parameters of the MM action (\ref{TEK-action}) as in 
(\ref{NN'Nk}),
block diagonal configurations, 
where the total number of blocks 
is specified by (\ref{blocknumberkcon}),
are dynamically favored.

We then consider small fluctuations around the minimum, i.e.,
configurations with $|q^a_l | \ll N_l$.
The classical action (\ref{clac}) is approximated as
\beq
\Delta S_b \simeq 4 \pi^2  \beta \frac{{\cal N}^2}{k} 
\sum_{l=1}^3 \frac{1}{(N_l)^4}\sum_{a=1}^h p^a (q^a_l)^2 \ ,
\label{deltaS}
\eeq
where we have given the difference from the minimum value.
We will call it an instanton action
since it is a classical action of a topological configuration.

For comparison, let us consider cases with large fluctuations,
for instance,
configurations where the total number of blocks is
different from (\ref{blocknumberkcon}),
and configurations specified by original tori with the integers $N_l$, $s_l$
that are different from the twists in the action (\ref{TEK-action}).
In these cases, the classical action (\ref{clac}) receives an enhancement factor
of order $(N_l)^2$, compared to (\ref{deltaS}).

\subsection{How to take large-$N$ limits and the probability distribution over the string vacuum space}
\label{sec:largeNprescription}

We now consider relations between the prescription of how to take a large-$N$ limit in the IIB MM
and the probability distribution over various matrix configurations, i.e.,
various string vacua,
based on the semiclassical analysis.
For details, see ref.~\cite{Aoki:2012ei}.

We first compare the IIB MM action (\ref{IIBMMaction}) 
and the unitary MM action (\ref{TEK-action}),
by considering a correspondence between the Hermitian matrices and the unitary matrices as
\beq
{\cal V}_M \sim \exp{\left(2 \pi i \frac{A_M}{\epsilon N_l}\right)} \ .
\label{corrHerUni}
\eeq
By plugging (\ref{corrHerUni}) into (\ref{TEK-action}),
and comparing it with (\ref{IIBMMaction}),
we find a relation among the coupling constants 
in (\ref{TEK-action}) and (\ref{IIBMMaction}) as
\beq
\frac{1}{2}\beta{\cal N}\left(\frac{2 \pi}{\epsilon N_l}\right)^4
=\frac{1}{g^2_{\rm IIBMM}} \ ,
\label{Corr_beta_gIIBMM}
\eeq
with $l=1,2,3$.
Similar relations can be obtained for $\beta'$ and $\beta''$.

Since both (\ref{deltaS}) and (\ref{Corr_beta_gIIBMM}) depend on $\beta/(N_l)^4$,
by defining a combination 
\beq
\frac{g^2_{\rm IIBMM}}{\epsilon^4 {\cal N}} \equiv \frac{1}{A} \ ,
\label{gepN-1}
\eeq
the instanton action (\ref{deltaS}) becomes 
\beq
\Delta S_b = \frac{A}{2 \pi^2 k}
\sum_{l=1}^3 \sum_{a=1}^h p^a (q^a_l)^2 \ .
\label{deltaST6dsl-1}
\eeq
It then follows that scaling limits of fixing
$
g^2_{\rm IIBMM}{\cal N}^\alpha/\epsilon^4
$
with $\alpha > -1$,
$\alpha = -1$,
and $\alpha < -1$
make the coefficient of the instanton action (\ref{deltaST6dsl-1})
infinite, finite, and vanishing values, respectively.
Together with fixing the torus size $\epsilon {\cal N}^{1/5}$,
those scaling limits correspond to 
fixing 
$
g^2_{\rm IIBMM}{\cal N}^{\gamma}
$
with 
$\gamma = \alpha +4/5$.
The three cases give drastically different results:

\begin{enumerate}
\item If we take a large-${\cal N}$ limit by fixing
$
g^2_{\rm IIBMM}{\cal N}^\alpha/\epsilon^4
$
with $\alpha> -1$, 
or by fixing 
$
g^2_{\rm IIBMM}{\cal N}^\gamma
$
with $\gamma > -1/5$, 
the instanton action (\ref{deltaST6dsl-1}) diverges for nonzero $q^a_l$,
and thus only a single topological sector survives. 
While in the present model, 
the topologically trivial sector,
$q^a_l=0$, is chosen, 
in more elaborate models,
desirable sectors may be chosen uniquely by the dynamics.
This is drastically different from the situations where
physicists usually consider the landscape.
\item In a limit with $\alpha< -1$ or $\gamma < -1/5$,
the instanton action (\ref{deltaST6dsl-1}) vanishes for all $q^a_l$,
and all the topological sectors appear with equal probabilities.
Then, the estimation for the probability distribution over the string vacuum space 
reduces to number counting of the classical solutions.
Moreover, in a limit with $\alpha < -1-2/5$,
a still larger number of configurations, 
which were studied as the large fluctuations below (\ref{deltaS}),
can also appear.
\item In a limit with $\alpha=-1$ or $\gamma = -1/5$,
the instanton action (\ref{deltaST6dsl-1}) takes finite values 
for finite $q^a_l$,
and all the topological sectors appear with finite
and different probabilities.
\end{enumerate}

\subsection{Probabilities for the appearance of the phenomenological models}

We now estimate the probabilities for the appearance
of the phenomenological models obtained in sections~\ref{sec:confpheno} and \ref{sec:conhiggs}.

We first consider the ${\rm U}(4) \times {\rm U}_L(2) \times {\rm U}_R(2)$ model
with the fluxes given by the first row in Table~\ref{table:q_ab_l_422}.
By solving (\ref{defqlab}), $q^a_l$ are determined as
\beqa
q^a_1 &=& (q_1,q_1+1,q_1-1) \ , \n
q^a_2 &=& (q_2,q_2-1,q_2-1) \ , \n
q^a_3 &=& (q_3,q_3-3,q_3-3) \ , 
\label{qla422_1}
\eeqa
for $a=1,\ldots, h=3$.
Since only the differences are specified in (\ref{defqlab}),
$q^a_l$ are determined with arbitrary integer shifts
$q_1$, $q_2$, and $q_3$.
The instanton action (\ref{deltaST6dsl-1}) takes various values
depending on these arbitrary integers $q_l$.
The minimum value 
\beq
\Delta S_b = \frac{A}{2 \pi^2 k} 28 \ ,
\label{MinInsAc422_1int}
\eeq
is attained by $q_1=0$, $q_2=0$ or 1, and $q_3=1$ or 2.  

One could further lower the instanton action (\ref{deltaST6dsl-1}) to
\beq
\Delta S_b = \frac{A}{2 \pi^2 k} 24 \ ,
\label{MinInsAc422_1fra}
\eeq
by choosing the fractional values $q_1=0$, $q_2=1/2$, and $q_3=3/2$.
This corresponds to
modifying the twists in the action (\ref{TEK-action})
from (\ref{twistsetting}) to
\beqa
{\cal Z}_{45}&=&
\exp{\left(2\pi i \frac{s_1}{N_1}\right)} \ , \n
{\cal Z}_{67}&=&
\exp{\left(2\pi i \left(\frac{s_2}{N_2}+\frac{1}{2 N_2^2}\right)\right)} \ , \n
{\cal Z}_{89}&=&
\exp{\left(2\pi i \left(\frac{s_3}{N_3}+\frac{1}{2 N_3^2}\right)\right)} \ ,
\label{twist422_1_relax}
\eeqa
although they are not natural in the present MM.

The probability of the appearance of this model
is semiclassically given as $e^{-\Delta S_b}$,
multiplied by a factor coming from quantum corrections.
There exist configurations with the same action,
but with $p^a$ and $q^a_l$ different from (\ref{qla422_1}),
and thus the probability must  
be divided by this numerical factor.

In the same way, we estimate the minimum instanton action for the other cases 
in Table~\ref{table:q_ab_l_422}.
We also examine the 
${\rm U}_c(3) \times {\rm U}(1) \times {\rm U}_L(2) \times {\rm U}_R(2)$ model
with the fluxes given in Table~\ref{table:q_ab_l_3122},
and the 
${\rm U}(4) \times {\rm U}_L(2) \times {\rm U}(1)^2$ model
with the fluxes in Table~\ref{table:q_ab_l_4211}.
The results are shown in Table~\ref{table:MinInsActU(8)}.
We further examine the models with the Higgsino fields, 
studied in section~\ref{sec:conhiggs},
and list the results in Table~\ref{table:MinInsActU(8)higgs}.
We do not present the results from the $T^6$ fluxes 
(\ref{q_ab_higgs3122_2}) or (\ref{q_ab_higgs31211_2}),
since they give larger values of the instanton actions.

\begin{table}
\begin{center}
\begin{tabular}{|c|c|c|}
\hline
gauge group & integral $q_l$ & fractional $q_l$  \\ \hline\hline
U(8) & 0 & 0 \\ \hline
${\rm U}(4) \times {\rm U}_L(2) \times {\rm U}_R(2)$ 
& 28 &24 \\
& 44 & 40 \\
& 36 & 32 \\
&36 & 32 \\ \hline
${\rm U}_c(3) \times {\rm U}(1) \times {\rm U}_L(2) \times {\rm U}_R(2)$ 
& 44 & 43 \\
&40&39\\
&36&36\\
&28&27 \\ \hline
${\rm U}(4) \times {\rm U}_L(2) \times {\rm U}(1)^2$ 
&28&25\\
&40&37\\
&32&29\\
&36&32\\
&44&40\\
&40&37\\
&36&36\\
&40&37\\
&36&33\\
\hline
\end{tabular} 
\caption{Values of the instanton actions 
$\frac{2 \pi^2 k}{A}  \cdot \Delta S_b$ in the
U(8) model,
the ${\rm U}(4) \times {\rm U}_L(2) \times {\rm U}_R(2)$ model,
the ${\rm U}_c(3) \times {\rm U}(1) \times {\rm U}_L(2) \times {\rm U}_R(2)$ model,
and the ${\rm U}(4) \times {\rm U}_L(2) \times {\rm U}(1)^2$ model,
without the Higgsino fields.
Each row of this table corresponds to 
those in Tables~\ref{table:q_ab_l_422},
\ref{table:q_ab_l_3122}, and
\ref{table:q_ab_l_4211}.
The second column shows the minimum value within the integer values of $q_l$
as in (\ref{MinInsAc422_1int}),
while the third column is the results within the fractional values of $q_l$
as in (\ref{MinInsAc422_1fra}).
Fractional values of $q_l$ are not natural in the present MM.
} 
\label{table:MinInsActU(8)}
\end{center}
\end{table}

\begin{table}
\begin{center}
\begin{tabular}{|c|c|c|c|}
\hline
gauge group & $q^{H_u}$, $q^{H_d}$ &integral $q_l$ & fractional $q_l$  \\ \hline\hline
${\rm U}(4) \times {\rm U}_L(2) \times {\rm U}_R(2)$ 
&24 & 44 &44 \\
&-8 & 28 & 28 \\
& 32 &44 & 40 \\
\hline
${\rm U}_c(3) \times {\rm U}(1) \times {\rm U}_L(2) \times {\rm U}_R(2)$ 
&-8 & 32 & 31 \\
&-8 &44&40\\
&-8&32&31\\
&32&44&43 \\ 
\hline
${\rm U}(4) \times {\rm U}_L(2) \times {\rm U}(1)^2$ 
&24, -8& 44&40\\
&-8, 32&40&39\\
&32, 24&44&43\\
&-8, -8&36&32\\
&32, 32&44&41\\
&0, 24& 40&39\\
&0, -8&28&27\\
&0, 32&36&36\\
&0, 24&44&43\\
&0, -8&36&35\\
&0, 32&40&37\\
&0, -8&32&31\\
&0, 32&44&41\\
\hline
${\rm U}_c(3) \times {\rm U}(1) \times {\rm U}_L(2) \times {\rm U}(1)^2$
&-8, -8&36&35\\
&0, -8&40&39\\
&0, 32&40&39\\
\hline
\end{tabular} 
\caption{Values of the instanton actions 
$\frac{2 \pi^2 k}{A}  \cdot \Delta S_b$ in the
the ${\rm U}(4) \times {\rm U}_L(2) \times {\rm U}_R(2)$ model,
the ${\rm U}_c(3) \times {\rm U}(1) \times {\rm U}_L(2) \times {\rm U}_R(2)$ model,
the ${\rm U}(4) \times {\rm U}_L(2) \times {\rm U}(1)^2$ model,
and the ${\rm U}_c(3) \times {\rm U}(1) \times {\rm U}_L(2) \times {\rm U}(1)^2$ model,
with the Higgsino candidates.
Each row of this table corresponds to those in Tables~\ref{table:q_ab_l_422higgs},
\ref{table:q_ab_l_3122higgs}, 
\ref{table:q_ab_l_4211higgs},
\ref{table:q_ab_l_4211higgs2}, and
\ref{table:q_ab_l_31211higgs}.
We do not present the results from the $T^6$ fluxes 
(\ref{q_ab_higgs3122_2}) or (\ref{q_ab_higgs31211_2}),
since they give larger values of the instanton actions.
The $T^6$ fluxes for the Higgsinos, $q^{H_u}$ and $q^{H_d}$, are 
also listed.
The third column shows the minimum value within the integer values of $q_l$
as in (\ref{MinInsAc422_1int}),
while the fourth column gives the results within the fractional values of $q_l$
as in (\ref{MinInsAc422_1fra}).
Fractional values of $q_l$ are not natural in the present MM.
} 
\label{table:MinInsActU(8)higgs}
\end{center}
\end{table}

Within these models and in the integral $q_l$ case,
the minimum instanton action $\frac{2 \pi^2 k}{A}  \cdot \Delta S_b$
takes various values between 28 and 44
over the various models.
The minimum of the minimum values, 28, is attained by
the first solution in the 
${\rm U}(4) \times {\rm U}_L(2) \times {\rm U}_R(2)$ model,
the last one in the 
${\rm U}_c(3) \times {\rm U}(1) \times {\rm U}_L(2) \times {\rm U}_R(2)$ model,
and the first one in the
${\rm U}(4) \times {\rm U}_L(2) \times {\rm U}(1)^2$ model
in Table~\ref{table:MinInsActU(8)},
and also by the second solution in the 
${\rm U}(4) \times {\rm U}_L(2) \times {\rm U}_R(2)$ model,
and the seventh one in the 
${\rm U}(4) \times {\rm U}_L(2) \times {\rm U}(1)^2$ model
in Table~\ref{table:MinInsActU(8)higgs}.
We cannot find drastic differences among the models with various gauge groups,
and between the models with and without the Higgsinos. 

There also exist solutions with lower instanton actions,
which do not yield the SM fermions.
Moreover, as far as we consider the models whose gauge group is a subgroup of U(8),
i.e., consider the MM (\ref{TEK-action}) with the parameter $k=8$,
the U(8) model without any gauge symmetry breaking has the lowest action.
In order to make phenomenologically attractive models most probable,
we need to elaborately modify the MM (\ref{TEK-action}).
However, we can also find that 
the phenomenological models have rather small instanton actions
and are sufficiently probable if the coefficient $A$ 
in the instanton action (\ref{deltaST6dsl-1})
is not so large.

\section{Conclusions and discussions}
\label{sec:conclusion}
\setcounter{equation}{0}

In this paper, we have considered the situations where the IIB MM is 
compactified on a torus with magnetic fluxes,
and exhausted all the matrix configurations that yield 
all the phenomenological models
whose gauge group is a subgroup of U(8),
with and without the Higgsino fields.
We have also studied the dynamics of MM semiclassically,
and estimated the probability distribution over the phenomenological models.

There remain some important problems.
While we found the embedding of the Higgs field in the matrices,
we need to consider how the electroweak symmetry breaking occurs.
The gauge-Higgs unifications and 
the recombinations of the intersecting D-branes
(see, for instance, ref.~\cite{Cremades:2002cs})
are close to the present situation
and may be helpful to consider this issue.
We should also study the values of the Yukawa couplings and 
the flavor structure.
They can be obtained by calculating overlaps among 
the zero-mode fields.
Related work has been done, for instance, in ref.~\cite{Abe:2012ya,Abe:2012fj,Cremades:2004wa}.

While we have considered situations where the supersymmetry 
protects the Higgs mass,
we need to study
how to keep part of the supersymmetry possessed by the IIB MM
and how to break it at low energies. 
It is also possible that the supersymmetry is broken at 
high energies but leaves some remnants
(see, for instance, ref.~\cite{Dienes:1995pm}).
Other resolutions of the naturalness or the hierarchy problem,
such as composite models, models with large extra dimensions,
some stringy or quantum-gravitational effects,
are also interesting to consider.
 
The models we have considered in the present paper have extra U(1) gauge groups
and are anomalous within the gauge dynamics.
This anomaly may be canceled via the Green-Schwarz mechanism
by the exchange of RR-fields,
which can also make the extra gauge fields massive.
In order to realize this, 
some constraints as the RR tadpole cancellation condition
should be imposed on MMs.
We should also generalize topological configurations
by introducing Wilson lines and tilting the tori.

While we assumed toroidal compactifications 
and worked in a unitary matrix formulation in this paper,
we should study the relation between the 
unitary MM and the IIB MM,
and how the parameters in the unitary MM action
are determined from the IIB MM.
We also need to consider how to interpret spacetime and matter in the matrices
and how to describe compactifications in the matrices.

We will come back to these issues in future publications.
We hope that these studies will give us some help for both 
exploring phenomenological models and
studying formulations of MMs.
We also would like to analyze the full dynamics of the MM,
and survey the probability distribution over the whole of the landscape.

\section*{Acknowledgements}
The author would like to thank S.~Iso, 
H.~Kawai, T.~Kobayashi, and J.~Nishimura
for valuable discussions.
This work is supported
in part by Grant-in-Aid for Scientific Research
(No. 24540279 and 23244057) from the Japan Society for the Promotion of Science.

\appendix

\section{Solutions of the $T^2$ fluxes $q^{ab}_l$}
\label{sec:q}
\setcounter{equation}{0}

In this appendix, we find solutions of the $T^2$ fluxes $q^{ab}_l$
by solving eq.~(\ref{defqab})
in the ${\rm U}_c(3) \times {\rm U}(1) \times {\rm U}_L(2) \times {\rm U}(1)^2$ model
and the 
${\rm U}_c(3) \times {\rm U}_L(2) \times {\rm U}(1)^4$ model
without the Higgsino fields,
studied in section~\ref{sec:confpheno}.

\subsection{Two theorems}

Before we solve eq.~(\ref{defqab})
 in the concrete models,
we prove two theorems.

{\bf Theorem~1}:
Pick up three elements from the $T^6$ fluxes
and focus on $q^{ab}$, $q^{bc}$, and $q^{ac}$:
\beq
\begin{pmatrix}
q^{ab} & q^{ac} \cr
& q^{bc}
\end{pmatrix} \ .
\eeq
Then, it is impossible that all of $|q^{ab}|$, $|q^{bc}|$, and $|q^{ac}|$
take 1, 2, or 3.
In other words, if $|q^{ab}|$, $|q^{bc}|$, and $|q^{ac}|$
are 0, 1, 2, or 3,
then, at least one of them must be 0.
There is one exception:
$(|q^{ab}|,|q^{bc}|,|q^{ac}|)=(2,2,2)$ is possible.

{\bf Proof}: 
Let us first consider the case $(|q^{ab}|,|q^{bc}|)=(1,1)$.
Then, the $T^2$ fluxes $|q^{ab}_l|$ must take 1 for all $l$.
So must $|q^{bc}_l|$.
It then follows from $q^{ac}_l = q^{ab}_l+q^{bc}_l$ that
$|q^{ac}_l|$ must be 0 or 2.
Then, $|q^{ac}|$ is 0 or 8.
Hence, $(|q^{ab}|,|q^{bc}|,|q^{ac}|)=(1,1,1)$,
$(1,1,2)$, or $(1,1,3)$
is not possible.

Similarly, by considering the cases $(|q^{ab}|,|q^{bc}|)=(1,2)$,
$(1,3)$, $(2,2)$, $(2,3)$, and $(3,3)$,
one can prove the theorem.

{\bf Theorem~2}: Consider the $T^6$ fluxes
\beq
q^{ab} = 
\begin{pmatrix}
0&0&-3&3\cr
&0&-3&x\cr
&&0&0\cr
&&&0\cr
\end{pmatrix} \ ,
\label{q_ab_333x}
\eeq
which is equivalent to
\beq
q^{ab} = 
\begin{pmatrix}
0&-3&0&3\cr
&0&3&0\cr
&&0&x\cr
&&&0\cr
\end{pmatrix} \ ,
\label{q_ab_333x2}
\eeq
as can be seen by exchanging the second and third rows and columns.
Then, it is impossible that $x$ takes $0$, $1$, $\pm 2$, or $-3$.
Within $|x| \le 3$, only $x=-1$ and $x=3$ are possible.

{\bf Proof}:
For 
\beq
q^{ab} = 
\begin{pmatrix}
0&-3&3\cr
&0&0\cr
&&0
\end{pmatrix} \ ,
\eeq
the solutions for eq.~(\ref{defqab}) are listed in 
Table \ref{table:q_ab_l_422}.
One can also list the solutions for
\beq
q^{ab} = 
\begin{pmatrix}
0&-3&x\cr
&0&0\cr
&&0
\end{pmatrix} \ .
\eeq
By combining two solutions from each,
one can construct a solution for (\ref{q_ab_333x}).
There is no solution for the cases $x=0, 1, \pm2, -3$.
For $x=3$, the solutions are listed in Table~\ref{table:q_ab_l_3122}.
For $x=-1$, there are three solutions,
which we list in Table~\ref{table:q_ab_l_333-1}.

\begin{table}
\begin{center}
\begin{tabular}{c|c|c}
$\hat{q}_1^{ab}$ & $\hat{q}_2^{ab}$ & $\hat{q}_3^{ab}$ \\ \hline\hline
$\begin{pmatrix}
-2&-1&-1\cr
&1&1\cr
&&0
\end{pmatrix}$
&
$\begin{pmatrix}
2&1&3\cr
&-1&1\cr
&&2
\end{pmatrix}$
&
$\begin{pmatrix}
0&3&-1\cr
&3&-1\cr
&&-4
\end{pmatrix}$
\\ \hline
$\begin{pmatrix}
0&-1&-1\cr
&-1&-1\cr
&&0
\end{pmatrix}$
&
$\begin{pmatrix}
-2&1&-3\cr
&3&-1\cr
&&-4
\end{pmatrix}$
&
$\begin{pmatrix}
2&3&1\cr
&1&-1\cr
&&-2
\end{pmatrix}$
\\ \hline
$\begin{pmatrix}
0&-1&-1\cr
&-1&-1\cr
&&0
\end{pmatrix}$
&
$\begin{pmatrix}
-2&1&-1\cr
&3&1\cr
&&-2
\end{pmatrix}$
&
$\begin{pmatrix}
2&3&3\cr
&1&1\cr
&&0
\end{pmatrix}$
\end{tabular} 
\caption{Fluxes in each $T^2$ for the $T^6$ fluxes (\ref{q_ab_333x})
with $x=-1$.
The diagonal elements are omitted.}
\label{table:q_ab_l_333-1}
\end{center}
\end{table}

\subsection{Fluxes in the 
${\rm U}_c(3) \times {\rm U}(1) \times {\rm U}_L(2) \times {\rm U}(1)^2$ model}
\label{sec:flux31211}

We now study 
the ${\rm U}_c(3) \times {\rm U}(1) \times {\rm U}_L(2) \times {\rm U}(1)^2$ model
considered in section \ref{sec:model31211}.
One can immediately find that 
(\ref{q_ab_31211II}) has no solution for (\ref{defqab}),
by applying Theorem~1 to
$q^{12}=3$, $q^{23}=-3$, and $q^{13}=-3$.

For (\ref{q_ab_31211I}),
by applying Theorem~1 to $q^{15}=3$,
one can find that 
$x=z=0$ or $x=z=3$ is allowed.
Also, by applying Theorem~1 to $q^{13}=-3$,
$x=y=0$ or $x=y=3$ is allowed.
Combining these two results,
$x=y=z=0$ or $x=y=z=3$ is allowed.
They correspond to
\beq
q^{ab}=
\begin{pmatrix}
0&0&-3&3&3\cr
&0&-3&\pm 3&3\cr
&&0&0&0\cr
&&&0&0\cr
&&&&0
\end{pmatrix} \ ,
\label{q_ab_31211I1}
\eeq
\beq
q^{ab}=
\begin{pmatrix}
0&3&-3&0&3\cr
&0&0&\mp 3&0\cr
&&0&3&0\cr
&&&0&3\cr
&&&&0
\end{pmatrix} \ ,
\label{q_ab_31211I2}
\eeq
respectively.
In fact, (\ref{q_ab_31211I1}) and (\ref{q_ab_31211I2}) are equivalent,
as one can see by exchanging the second and fourth rows and columns,
i.e., by exchanging the first and second U(1).

By applying Theorem~2 to the first $4 \times 4$ part of (\ref{q_ab_31211I1}),
one can find that the plus sign in the double sign must be chosen.
Then, the problem comes back to the case in (\ref{q_ab_31211}).
Thus, we have to conclude again that there is no solution in the
${\rm U}_c(3) \times {\rm U}(1) \times {\rm U}_L(2) \times {\rm U}(1)^2$
model, even though the fermion embedding is generalized to
(\ref{fermionembed31211I}) and (\ref{fermionembed31211II}). 

\subsection{Fluxes in the ${\rm U}_c(3) \times {\rm U}_L(2) \times {\rm U}(1)^4$ model}
\label{sec:flux321111}

We then study 
the ${\rm U}_c(3) \times {\rm U}_L(2) \times {\rm U}(1)^4$ model
considered in section \ref{sec:model321111}.
One can immediately find that the case 
(\ref{fermionembed321111III}) has no solution for $q^{ab}_l$,
by applying Theorem~1 to
$q^{12}=-3$, $q^{23}=3$, and $q^{13}=3$.

For the case (\ref{fermionembed321111II}),
by using Theorem~1,
the possible $T^6$ fluxes 
turn out to be as follows:
\beq
q^{ab}=
\begin{pmatrix}
0&-3&0&3&1&2\cr
&0&3&0&0&0\cr
&&0&\pm 3&x&3-x\cr
&&&0&0&0\cr
&&&&0&0\cr
&&&&&0
\end{pmatrix} \ ,
\label{q^ab321111II1}
\eeq
\beq
q^{ab}=
\begin{pmatrix}
0&-3&0&3&3&0\cr
&0&3&0&0&0\cr
&&0&u&x&y\cr
&&&0&0&z\cr
&&&&0&v\cr
&&&&&0
\end{pmatrix} \ ,\label{q^ab321111II2}
\eeq
where $x$, $y$, and $z$ can take an integer 0, 1, 2, or 3,
while $u$ and $v$ can take 0, $\pm 1$, $\pm 2$, or $\pm 3$.
In (\ref{q^ab321111II2}), $x+y+z=3$ and $|u|+|v|=3$
must be satisfied.

For (\ref{q^ab321111II1}), 
by applying Theorem~2 to
the first $4 \times 4$ part, i.e.,
$q^{ab}$ with $1 \le a,b \le 4$,
one can see that the plus sign in the double sign must be chosen.
Then, $q^{ab}$ with $1 \le a,b \le 4$
are the same as (\ref{q_ab_3122}),
as one can see by exchanging the second and third rows and columns 
in $q^{ab}$.
Thus, we can construct solutions for (\ref{q^ab321111II1})
by using the solutions in Table \ref{table:q_ab_l_3122}.
There is one solution.
The $T^2$ fluxes are given 
in the first row in Table \ref{table:q_ab_l_321111II}.
The corresponding $T^6$ fluxes are (\ref{q^ab321111II1}) with $x=1$, 
i.e.,
\beq
q^{ab}=
\begin{pmatrix}
0&-3&0&3&1&2 \cr
&0&3&0&0&0\cr
&&0&3&1&2\cr
&&&0&0&0\cr
&&&&0&0\cr
&&&&&0
\end{pmatrix} \ .
\label{q_ab_321111II1}
\eeq

For (\ref{q^ab321111II2}),
by applying Theorem~2 to the first $4 \times 4$ part, i.e.,
$q^{ab}$ with $1 \le a,b \le 4$,
one can find that $u$ must take $-1$ or $3$.
In the same way, $x$ must be $-1$ or $3$.
As we showed below (\ref{q_ab_31211}),
there is no solution for $(u,x)=(3,3)$.
Since $x$ must be positive, the remaining possibility is
$(u,x)=(-1,3)$.
It then follows from $x+y+z=3$ and $|u|+|v|=3$ that
$y=z=0$ and $|v|=2$.
One can construct solutions by using 
those in Tables~\ref{table:q_ab_l_3122}
and \ref{table:q_ab_l_333-1}.
There is one solution.
The $T^2$ fluxes are given in the second row
in Table \ref{table:q_ab_l_321111II}.
The corresponding $T^6$ fluxes are 
\beq
q^{ab}=
\begin{pmatrix}
0&-3&0&3&3&0 \cr
&0&3&0&0&0\cr
&&0&-1&3&0\cr
&&&0&0&0\cr
&&&&0&-2\cr
&&&&&0
\end{pmatrix} \ .
\label{q_ab_321111II2}
\eeq

\begin{table}
\begin{center}
\begin{tabular}{c|c|c}
$\hat{q}_1^{ab}$ & $\hat{q}_2^{ab}$ & $\hat{q}_3^{ab}$ \\ \hline\hline
$\begin{pmatrix}
-1&-2&-1&-1&-1\cr
&-1&0&0&0\cr
&&1&1&1\cr
&&&0&0\cr
&&&&0
\end{pmatrix}$
&
$\begin{pmatrix}
-1&-2&-1&-1&-1\cr
&-1&0&0&0\cr
&&1&1&1\cr
&&&0&0\cr
&&&&0
\end{pmatrix}$
&
$\begin{pmatrix}
-3&0&3&1&2\cr
&3&6&4&5\cr
&&3&1&2\cr
&&&-2&-1\cr
&&&&1
\end{pmatrix}$
\\ \hline
$\begin{pmatrix}
-1&0&-1&1&-1\cr
&1&0&2&0\cr
&&-1&1&-1\cr
&&&2&0\cr
&&&&-2
\end{pmatrix}$
&
$\begin{pmatrix}
1&-2&-1&1&0\cr
&-3&-2&0&-1\cr
&&1&3&2\cr
&&&2&1\cr
&&&&-1
\end{pmatrix}$
&
$\begin{pmatrix}
3&2&3&3&2\cr
&-1&0&0&-1\cr
&&1&1&0\cr
&&&0&-1\cr
&&&&-1
\end{pmatrix}$
\end{tabular} 
\caption{Fluxes in each $T^2$ in the 
${\rm U}_c(3) \times {\rm U}_L(2) \times {\rm U}(1)^4$ model
with the fermion embedding (\ref{fermionembed321111II}).
The first and second rows of this table
correspond to the $T^6$ fluxes 
(\ref{q_ab_321111II1}) and (\ref{q_ab_321111II2}),
respectively.
The diagonal elements are omitted.}
\label{table:q_ab_l_321111II}
\end{center}
\end{table}

For the case (\ref{fermionembed321111I}),
by using Theorem~1,
the possible $T^6$ fluxes 
turn out to be as follows:
\beq
q^{ab}=
\begin{pmatrix}
0&-3&0&1&2&3\cr
&0&3&0&0&0\cr
&&0&u &v &3\cr
&&&0&0&0\cr
&&&&0&0\cr
&&&&&0
\end{pmatrix} \ ,
\label{q^ab321111I1}
\eeq
\beq
q^{ab}=
\begin{pmatrix}
0&-3&0&0&3&3\cr
&0&1&2&0&0\cr
&&0&0&u&x\cr
&&&0&v&3-x\cr
&&&&0&0\cr
&&&&&0
\end{pmatrix} \ ,
\label{q^ab321111I2}
\eeq
\beq
q^{ab}=
\begin{pmatrix}
0&-3&0&3&0&3\cr
&0&3&0&0&0\cr
&&0&u &v&x\cr
&&&0&w&0\cr
&&&&0&3-x\cr
&&&&&0
\end{pmatrix} \ ,
\label{q^ab321111I3}
\eeq
where $x$ can take an integer  0, 1, 2, or 3,
while $u$, $v$, and $w$ can take 0, $\pm 1$, $\pm 2$, or $\pm 3$.
In (\ref{q^ab321111I1}) and (\ref{q^ab321111I2}),
$|u|+|v|=3$ is required.
In (\ref{q^ab321111I3}),
$|u|+|v|+|w|=3$ is required.

Since (\ref{q^ab321111I1}) has a similar structure to 
(\ref{q^ab321111II1}), as can be seen by permuting the last 
three columns, one can find solutions in the same way.
There are three solutions:
The $T^2$ fluxes of the first row in Table \ref{table:q_ab_l_321111I1}
yield the $T^6$ fluxes (\ref{q_ab_321111I1}).
The second and third rows in Table \ref{table:q_ab_l_321111I1}
give (\ref{q_ab_321111I2}).
\beq
q^{ab}=
\begin{pmatrix}
0&-3&0&1&2&3 \cr
&0&3&0&0&0\cr
&&0&1&2&3\cr
&&&0&0&0\cr
&&&&0&0\cr
&&&&&0
\end{pmatrix} 
\label{q_ab_321111I1}
\eeq
\beq
q^{ab}=
\begin{pmatrix}
0&-3&0&1&2&3 \cr
&0&3&0&0&0\cr
&&0&-3&0&3\cr
&&&0&0&0\cr
&&&&0&0\cr
&&&&&0
\end{pmatrix} 
\label{q_ab_321111I2}
\eeq

For (\ref{q^ab321111I2}), $q^{ab}$ with $a, b=1,2,5,6$ 
 are the same as (\ref{q_ab_4211}),
and thus one can use the results in Table \ref{table:q_ab_l_4211}.
We simply find solutions for the first $4 \times 4$ part
in (\ref{q^ab321111I2}).
We then combine the results,
and obtain solutions for (\ref{q^ab321111I2}).
There are three solutions:
The $T^2$ fluxes of the fourth and fifth 
rows in Table~\ref{table:q_ab_l_321111I1}
yield $T^6$ fluxes (\ref{q_ab_321111I3}).
The sixth row in Table~\ref{table:q_ab_l_321111I1}
gives (\ref{q_ab_321111I4}).
\beq
q^{ab}=
\begin{pmatrix}
0&-3&0&0&3&3 \cr
&0&1&2&0&0\cr
&&0&0&-3&1\cr
&&&0&0&2\cr
&&&&0&0\cr
&&&&&0
\end{pmatrix} 
\label{q_ab_321111I3}
\eeq
\beq
q^{ab}=
\begin{pmatrix}
0&-3&0&0&3&3 \cr
&0&1&2&0&0\cr
&&0&0&1&1\cr
&&&0&2&2\cr
&&&&0&0\cr
&&&&&0
\end{pmatrix} 
\label{q_ab_321111I4}
\eeq

For (\ref{q^ab321111I3}),
$q^{ab}$ with $a,b=1,2,3,4,6$ are the same as
the first $5 \times 5$ part in (\ref{q^ab321111II2}).
One can thus follow the same arguments there.
There are nine solutions:
The $T^2$ fluxes in
Table~\ref{table:q_ab_l_321111I2}
yield the $T^6$ fluxes (\ref{q_ab_321111I5}),
and those in Table~\ref{table:q_ab_l_321111I3}
give (\ref{q_ab_321111I6}).
\beq
q^{ab}=
\begin{pmatrix}
0&-3&0&3&0&3 \cr
&0&3&0&0&0\cr
&&0&-1&\pm 2&3\cr
&&&0&0&0\cr
&&&&0&0\cr
&&&&&0
\end{pmatrix} 
\label{q_ab_321111I5}
\eeq
\beq
q^{ab}=
\begin{pmatrix}
0&-3&0&3&0&3 \cr
&0&3&0&0&0\cr
&&0&-1&0&3\cr
&&&0&\pm 2&0\cr
&&&&0&0\cr
&&&&&0
\end{pmatrix} 
\label{q_ab_321111I6}
\eeq

In all of the above solutions (\ref{q_ab_321111II1}), (\ref{q_ab_321111II2}),
(\ref{q_ab_321111I1})-(\ref{q_ab_321111I6}), some of the fermion species are 
placed in several matrix elements, i.e.,
not all of the fermion species 
can be put in a single place with three generations.
This is inevitable since there is no solution for (\ref{q_ab_31211}).
This may give some interesting results in the flavor structure.

\begin{table}
\begin{center}
\begin{tabular}{c|c|c}
$\hat{q}_1^{ab}$ & $\hat{q}_2^{ab}$ & $\hat{q}_3^{ab}$ \\ \hline\hline
$\begin{pmatrix}
-1&-2&-1&-1&-1\cr
&-1&0&0&0\cr
&&1&1&1\cr
&&&0&0\cr
&&&&0
\end{pmatrix}$
&
$\begin{pmatrix}
-1&-2&-1&-1&-1\cr
&-1&0&0&0\cr
&&1&1&1\cr
&&&0&0\cr
&&&&0
\end{pmatrix}$
&
$\begin{pmatrix}
-3&0&1&2&3\cr
&3&4&5&6\cr
&&1&2&3\cr
&&&1&2\cr
&&&&1
\end{pmatrix}$
\\ \hline
$\begin{pmatrix}
-1&-2&-1&-1&-1\cr
&-1&0&0&0\cr
&&1&1&1\cr
&&&0&0\cr
&&&&0
\end{pmatrix}$
&
$\begin{pmatrix}
-1&-2&1&-2&-1\cr
&-1&2&-1&0\cr
&&3&0&1\cr
&&&-3&-2\cr
&&&&1
\end{pmatrix}$
&
$\begin{pmatrix}
-3&0&-1&1&3\cr
&3&2&4&6\cr
&&-1&1&3\cr
&&&2&4\cr
&&&&2
\end{pmatrix}$
\\ \hline
$\begin{pmatrix}
1&0&-1&-1&-1\cr
&-1&-2&-2&-2\cr
&&-1&-1&-1\cr
&&&0&0\cr
&&&&0
\end{pmatrix}$
&
$\begin{pmatrix}
3&2&1&2&3\cr
&-1&-2&-1&0\cr
&&-1&0&1\cr
&&&1&2\cr
&&&&1
\end{pmatrix}$
&
$\begin{pmatrix}
-1&2&-1&-1&-1\cr
&3&0&0&0\cr
&&-3&-3&-3\cr
&&&0&0\cr
&&&&0
\end{pmatrix}$
\\ \hline
$\begin{pmatrix}
-1&0&0&-1&1\cr
&1&1&0&2\cr
&&0&-1&1\cr
&&&-1&1\cr
&&&&2
\end{pmatrix}$
&
$\begin{pmatrix}
1&2&-1&-1&1\cr
&1&-2&-2&0\cr
&&-3&-3&-1\cr
&&&0&2\cr
&&&&2
\end{pmatrix}$
&
$\begin{pmatrix}
3&4&2&3&3\cr
&1&-1&0&0\cr
&&-2&-1&-1\cr
&&&1&1\cr
&&&&0
\end{pmatrix}$
\\ \hline
$\begin{pmatrix}
-1&0&0&3&1\cr
&1&1&4&2\cr
&&0&3&1\cr
&&&3&1\cr
&&&&-2
\end{pmatrix}$
&
$\begin{pmatrix}
1&0&0&1&1\cr
&-1&-1&0&0\cr
&&0&1&1\cr
&&&1&1\cr
&&&&0
\end{pmatrix}$
&
$\begin{pmatrix}
3&2&1&1&3\cr
&-1&-2&-2&0\cr
&&-1&-1&1\cr
&&&0&2\cr
&&&&2
\end{pmatrix}$
\\ \hline
$\begin{pmatrix}
-1&0&0&1&-1\cr
&1&1&2&0\cr
&&0&1&-1\cr
&&&1&-1\cr
&&&&-2
\end{pmatrix}$
&
$\begin{pmatrix}
1&0&0&1&-1\cr
&-1&-1&0&-2\cr
&&0&1&-1\cr
&&&1&-1\cr
&&&&-2
\end{pmatrix}$
&
$\begin{pmatrix}
3&2&1&3&3\cr
&-1&-2&0&0\cr
&&-1&1&1\cr
&&&2&2\cr
&&&&0
\end{pmatrix}$
\end{tabular} 
\caption{Fluxes in each $T^2$ in the ${\rm U}_c(3) \times {\rm U}_L(2) \times {\rm U}(1)^4$ model
with the fermion embedding (\ref{fermionembed321111I}) (part~1).
The first row of this table corresponds to the $T^6$ fluxes (\ref{q_ab_321111I1}),
the second and third rows to (\ref{q_ab_321111I2}),
the fourth and fifth rows to (\ref{q_ab_321111I3}),
and the sixth row to (\ref{q_ab_321111I4}).
The diagonal elements are omitted.}
\label{table:q_ab_l_321111I1}
\end{center}
\end{table}

\begin{table}
\begin{center}
\begin{tabular}{c|c|c}
$\hat{q}_1^{ab}$ & $\hat{q}_2^{ab}$ & $\hat{q}_3^{ab}$ \\ \hline\hline
$\begin{pmatrix}
-1&-2&-1&0&-1\cr
&-1&0&1&0\cr
&&1&2&1\cr
&&&1&0\cr
&&&&-1
\end{pmatrix}$
&
$\begin{pmatrix}
1&2&3&1&1\cr
&1&2&0&0\cr
&&1&-1&-1\cr
&&&-2&-2\cr
&&&&0
\end{pmatrix}$
&
$\begin{pmatrix}
3&0&-1&-1&-3\cr
&-3&-4&-4&-6\cr
&&-1&-1&-3\cr
&&&0&-2\cr
&&&&-2
\end{pmatrix}$
\\ \hline 
$\begin{pmatrix}
-1&-2&-1&-1&-1\cr
&-1&0&0&0\cr
&&1&1&1\cr
&&&0&0\cr
&&&&0
\end{pmatrix}$
&
$\begin{pmatrix}
1&2&3&0&1\cr
&1&2&-1&0\cr
&&1&-2&-1\cr
&&&-3&-2\cr
&&&&1
\end{pmatrix}$
&
$\begin{pmatrix}
3&0&-1&-1&-3\cr
&-3&-4&-4&-6\cr
&&-1&-1&-3\cr
&&&0&-2\cr
&&&&-2
\end{pmatrix}$
\\ \hline 
$\begin{pmatrix}
-1&-2&-1&-1&-1\cr
&-1&0&0&0\cr
&&1&1&1\cr
&&&0&0\cr
&&&&0
\end{pmatrix}$
&
$\begin{pmatrix}
1&2&3&0&1\cr
&1&2&-1&0\cr
&&1&-2&-1\cr
&&&-3&-2\cr
&&&&1
\end{pmatrix}$
&
$\begin{pmatrix}
3&0&-1&1&-3\cr
&-3&-4&-2&-6\cr
&&-1&1&-3\cr
&&&2&-2\cr
&&&&-4
\end{pmatrix}$
\\ \hline 
$\begin{pmatrix}
-1&0&-1&1&1\cr
&1&0&2&2\cr
&&-1&1&1\cr
&&&2&2\cr
&&&&0
\end{pmatrix}$
&
$\begin{pmatrix}
1&-2&-1&0&1\cr
&-3&-2&-1&0\cr
&&1&2&3\cr
&&&1&2\cr
&&&&1
\end{pmatrix}$
&
$\begin{pmatrix}
3&2&3&3&3\cr
&-1&0&0&0\cr
&&1&1&1\cr
&&&0&0\cr
&&&&0
\end{pmatrix}$
\\ \hline 
$\begin{pmatrix}
-1&0&-1&-1&1\cr
&1&0&0&2\cr
&&-1&-1&1\cr
&&&0&2\cr
&&&&2
\end{pmatrix}$
&
$\begin{pmatrix}
1&-2&-1&0&1\cr
&-3&-2&-1&0\cr
&&1&2&3\cr
&&&1&2\cr
&&&&1
\end{pmatrix}$
&
$\begin{pmatrix}
3&2&3&3&3\cr
&-1&0&0&0\cr
&&1&1&1\cr
&&&0&0\cr
&&&&0
\end{pmatrix}$
\end{tabular} 
\caption{Fluxes in each $T^2$ in the ${\rm U}_c(3) \times {\rm U}_L(2) \times {\rm U}(1)^4$ model
with the fermion embedding (\ref{fermionembed321111I})
(part~2).
The solutions in this table correspond to
the $T^6$ fluxes (\ref{q_ab_321111I5}).
The diagonal elements are omitted.}
\label{table:q_ab_l_321111I2}
\end{center}
\end{table}

\begin{table}
\begin{center}
\begin{tabular}{c|c|c}
$\hat{q}_1^{ab}$ & $\hat{q}_2^{ab}$ & $\hat{q}_3^{ab}$ \\ \hline\hline
$\begin{pmatrix}
-1&-2&-1&-2&-1\cr
&-1&0&-1&0\cr
&&1&0&1\cr
&&&-1&0\cr
&&&&1
\end{pmatrix}$
&
$\begin{pmatrix}
1&2&3&1&1\cr
&1&2&0&0\cr
&&1&-1&-1\cr
&&&-2&-2\cr
&&&&0
\end{pmatrix}$
&
$\begin{pmatrix}
3&0&-1&0&-3\cr
&-3&-4&-3&-6\cr
&&-1&0&-3\cr
&&&1&-2\cr
&&&&-3
\end{pmatrix}$
\\ \hline 
$\begin{pmatrix}
-1&-2&-1&0&-1\cr
&-1&0&1&0\cr
&&1&2&1\cr
&&&1&0\cr
&&&&1
\end{pmatrix}$
&
$\begin{pmatrix}
1&2&3&1&1\cr
&1&2&0&0\cr
&&1&-1&-1\cr
&&&-2&-2\cr
&&&&0
\end{pmatrix}$
&
$\begin{pmatrix}
3&0&-1&0&-3\cr
&-3&-4&-3&-6\cr
&&-1&0&-3\cr
&&&1&-2\cr
&&&&-3
\end{pmatrix}$
\\ \hline 
$\begin{pmatrix}
-1&0&-1&0&1\cr
&1&0&1&2\cr
&&-1&0&1\cr
&&&1&2\cr
&&&&1
\end{pmatrix}$
&
$\begin{pmatrix}
1&-2&-1&1&1\cr
&-3&-2&0&0\cr
&&1&3&3\cr
&&&2&2\cr
&&&&0
\end{pmatrix}$
&
$\begin{pmatrix}
3&2&3&4&3\cr
&-1&0&1&0\cr
&&1&2&1\cr
&&&1&0\cr
&&&&-1
\end{pmatrix}$
\\ \hline 
$\begin{pmatrix}
-1&0&-1&0&1\cr
&1&0&1&2\cr
&&-1&0&1\cr
&&&1&2\cr
&&&&1
\end{pmatrix}$
&
$\begin{pmatrix}
1&-2&-1&1&1\cr
&-3&-2&0&0\cr
&&1&3&3\cr
&&&2&2\cr
&&&&0
\end{pmatrix}$
&
$\begin{pmatrix}
3&2&3&2&3\cr
&-1&0&-1&0\cr
&&1&0&1\cr
&&&-1&0\cr
&&&&1
\end{pmatrix}$
\end{tabular} 
\caption{Fluxes in each $T^2$ in the ${\rm U}_c(3) \times {\rm U}_L(2) \times {\rm U}(1)^4$ model
with the fermion embedding (\ref{fermionembed321111I})
(part~3).
The solutions in this table correspond to
the $T^6$ fluxes (\ref{q_ab_321111I6}).
The diagonal elements are omitted.}
\label{table:q_ab_l_321111I3}
\end{center}
\end{table}

\end{document}